\documentclass[10pt,journal,final,twocolumn]{IEEEtran}
\usepackage{amsmath}
\usepackage{epsfig,graphicx,subfigure,psfrag,amsmath,cases,bm}
\usepackage{latexsym,amssymb,algorithm,mathtools}
\usepackage{algorithmic}
\usepackage{color}
\usepackage{url}
\usepackage{scrtime}
\usepackage{stfloats}
\usepackage{tablefootnote}
\usepackage{cite}
\usepackage{amsfonts,mathabx}
\usepackage{textcomp}
\usepackage{caption}
\usepackage{microtype}
\usepackage{xcolor,capt-of}
\usepackage{multirow}
\usepackage{balance}
\usepackage{xcolor}
\usepackage{hyperref}
\newcommand\scalemath[2]{\scalebox{#1}{\mbox{\ensuremath{\displaystyle #2}}}}
\begin{document}
\newtheorem{lemma}{Lemma}
    \title{\huge Efficient UAV Hovering, Resource Allocation, and Trajectory Design for ISAC with Limited Backhaul Capacity}
\author{ Ata Khalili, \textit{Member, IEEE}, Atefeh Rezaei, \textit{Student Member, IEEE}, Dongfang Xu, \textit{Member, IEEE}, Falko Dressler, \textit{Fellow, IEEE}, and Robert Schober, \textit{Fellow, IEEE} 
\thanks{This work was supported partly by the Federal Ministry of Education and Research of Germany under the program of “Souveran. Digital. Vernetzt.” joint project 6G-RIC (project identification number: PIN 16KISK023) and also in part by the Deutsche Forschungsgemeinschaft (DFG, German Research Foundation) GRK-2680 – Project-ID 437847244. \\ Part of this paper has been presented at the IEEE Global Communication Conference (Globecom) 2023 \cite{Globecom2023}. A. Khalili and R. Schober are with the Institute for Digital Communications, Friedrich- Alexander-University Erlangen–Nurnberg, 91054 Erlangen, Germany (e-mail: ata.khalili@fau.de, robert.schober@fau.de). A. Rezaei and F. Dressler are with the School of Electrical Engineering and
Computer Science, Technische Universität Berlin, Germany (e-mail: rezaei@ccs-labs.org, dressler@ccs-labs.org). D. Xu is with the Hong Kong University of Science and Technology, Hong Kong (e-mail: eedxu@ust.hk).}}


\maketitle

\begin{abstract}
In this paper, we investigate the joint resource allocation and trajectory design for a multi-user, multi-target unmanned aerial vehicle (UAV)-enabled integrated sensing and communication (ISAC) system, where the link capacity between a ground base station (BS) and the UAV is limited. The UAV conducts target sensing and information transmission in orthogonal time slots to prevent interference. As is common in practical systems, sensing is performed while the UAV hovers, allowing the UAV to acquire high-quality sensing data. Subsequently, the acquired sensing data is offloaded to the ground BS for further processing. We jointly optimize the UAV trajectory, UAV velocity, beamforming for the communication users, power allocated to the sensing beam, and time of hovering for sensing to minimize the power consumption of the UAV while ensuring the communication quality of service (QoS) and successful sensing. Due to the prohibitively high complexity of the resulting non-convex mixed integer non-linear program (MINLP), we employ a series of transformations and optimization techniques, including semidefinite relaxation, big-M method, penalty approach, and successive convex approximation, to obtain a low-complexity suboptimal solution. 
Our simulation results reveal that 1) the proposed design achieves significant power savings compared to two baseline schemes; 2) stricter sensing requirements lead to longer sensing times, highlighting the challenge of efficiently managing both sensing accuracy and sensing time; 3) the optimized trajectory design ensures precise hovering directly above the targets during sensing, enhancing sensing quality and enabling the application of energy-focused beams; and 4) the proposed trajectory design balances the capacity of the backhaul link and the downlink rate of the communication users.


\end{abstract}

\vspace{-8mm}
\section{Introduction}
With the rapid development of new services for future wireless networks, the sixth generation (6G) of wireless communication systems is expected to become a fully intelligent network enabling a multitude of environment-and location-aware applications such as autonomous driving, remote healthcare, and smart industry. To support these applications, a seamless 6G wireless network is needed, providing both high-precision sensing capabilities and wireless information transmission. To this end, integrated sensing and communication (ISAC) has recently drawn significant attention from academia and industry. ISAC is capable of increasing spectrum efficiency and facilitating the sharing of the physical infrastructure for sensing and communications \cite{ISAC6G}. 
Motivated by these advantages, the authors in \cite{mu-mimo-jsc,jsc-mimo-radar} considered ISAC networks with a specific focus on terrestrial systems. However, terrestrial ISAC systems, despite their potential, are often hindered by obstacles on the ground that may obstruct the line of sight (LoS) to sensing targets.

Compared to conventional cellular systems which are based on a fixed terrestrial infrastructure, unmanned aerial vehicle (UAV)-enabled communication systems can support on-demand connectivity by flexibly deploying UAV-enabled wireless transceivers in a target area. For example, in the case of natural disasters and major accidents, UAVs can be utilized as aerial base stations to establish temporary communication links in a timely and cost-effective manner. Moreover, UAV-aided wireless communication, capitalizing on the flexible deployment\cite{SurveyUAV,DUAV}, can exploit LoS links. These LoS connections cannot only enhance communication performance but also serve as a critical element for accurate target sensing. This is because target detection and parameter estimation usually require LoS links between the sensing transceivers and the sensing targets. Furthermore, due to their high maneuverability, UAVs can quickly approach a desired target, which can significantly reduce the transmit power required for sensing\cite{CSUAV,Mag_Qing}. Despite these promising features, only few works in the existing literature have studied UAV-enabled ISAC \cite{maneuver,UAVISAC1,ThUAVISAC,UAVISACSecure,UAVLiu}. The authors in \cite{maneuver} optimized the trajectory, transmit beamforming, and radar signals of a UAV-enabled ISAC system to improve the communication data rate while ensuring a required sensing beam pattern gain. In \cite{UAVISAC1,ThUAVISAC}, a periodic sensing and communication scheme for UAV-enabled ISAC systems was introduced and the achievable rate was maximized by jointly optimizing the UAV's trajectory, transmit precoder, and sensing start time subject to sensing frequency and beam pattern gain constraints. The authors in \cite{UAVISACSecure} proposed a novel integrated sensing, jamming, and communication framework for UAV-enabled downlink communications to maximize the number of securely served users while considering a tracking performance constraint. In \cite{UAVLiu}, the authors considered single-antenna UAV-enabled integrated sensing, computing, and communication, where the UAV sensed a target and offloaded computational tasks to the ground base station (BS). 

 Despite the comprehensiveness of the studies in  \cite{UAVISAC1,ThUAVISAC,UAVISACSecure,UAVLiu}, they did not consider the aerodynamic power consumption and velocity optimization of the UAV, which is crucial for overcoming the limited battery capacity of UAVs. Efficient and prolonged UAV operation requires addressing these aspects via resource allocation and trajectory design, enhancing overall performance and mission time. 
Furthermore, the authors of \cite{maneuver,UAVISAC1,ThUAVISAC,UAVISACSecure,UAVLiu} primarily concentrated on optimizing the beam pattern gain for target sensing, while ignoring the potential impact of the sidelobes of the beam pattern. In fact, the presence of sidelobes can result in energy wastage and cause interference, which may have adverse effects on the overall performance of ISAC systems\cite{jsc-mimo-radar,mu-mimo-jsc}. Moreover, these studies did not consider the signal-to-noise ratio (SNR) of the received radar echoes as a performance metric for sensing. However, the reliable detection of radar echoes is essential for successful sensing in practice. Therefore, in this paper, we consider the SNR of the received radar echoes as a performance metric for sensing. Besides, the ISAC systems considered in \cite{Globecom2023,UAVISAC1,ThUAVISAC,UAVISACSecure,UAVLiu} may experience significant self-interference (SI) as the radar echoes may be received while data is being transmitted. Although conventional full-duplex communication systems use SI cancellation techniques to mitigate such interference, these methods may not be sufficient to suppress the SI below the level required for sensing due to the low received echo powers. This is primarily due to the high attenuation of the echo signal caused by the round-trip path-loss, which makes it challenging to achieve sufficient SI suppression. To address this issue, we propose to perform sensing and communication in orthogonal time slots, coupled with the adoption of pulse radar technology, which enables flexible adjustment of the sensing range \cite{Ata_ICC}. Besides, in \cite{maneuver,UAVISAC1,ThUAVISAC,UAVISACSecure,UAVLiu}, sensing was performed while the UAV was moving, which can potentially degrade sensing accuracy. In contrast, practical UAV-based sensing systems typically perform sensing while the UAV is hovering\cite{UAVr}. Therefore, in this paper, we incorporate this feature into our problem formulation to leverage the following advantages. Firstly, when the UAV hovers above the target, a predetermined fixed beam pattern can be used for sensing. This eliminates the need for continuous adjustment of the beam pattern based on the UAV's flight path, which significantly reduces design complexity. Secondly, hovering during sensing circumvents the UAV-induced Doppler shift, simplifying sensing data signal processing. Thirdly, this positioning strategy also ensures a direct LoS link to the target, which is crucial for reliable sensing. Furthermore, in \cite{Globecom2023,UAVISAC1,ThUAVISAC,UAVISACSecure}, it was assumed that the processing of the received sensing signals is done locally at the UAV, which can be challenging due to the UAV's limited computational capabilities and battery resources. In fact, signal processing and analysis of sensing data require significant computing resources and energy, which creates a bottleneck for UAV-enabled sensing. To address this challenge, we propose to forward the received sensing echo signals to a ground BS via backhaul links for processing. Offloading the sensing data alleviates the computational burden for the UAV, enabling higher accuracy and lower latency in obtaining sensing results at the BS. 

In summary existing studies did not account for critical aspects of practical ISAC systems such as aerodynamic power consumption, velocity optimization, received echo SNR as performance metric for sensing, synthesizing a focused beam, hovering during sensing, and the impact of a limited backhaul link capacity on system performance. This paper addresses these critical aspects and introduces a novel design framework for minimization of the average UAV power consumption while meeting the quality of service (QoS) requirements of both the communication users and the sensing tasks.
The main contributions of this paper can be summarized as follows:
	\begin{itemize}
	    \item We consider a multi-user multi-target UAV-enabled ISAC system where the link capacity between the ground BS and the UAV is limited. We aim to minimize the average power consumption of the UAV, which involves optimizing not only the resource allocation and UAV trajectory but also the time when the UAV hovers for sensing, resulting in a non-convex mixed integer non-linear program (MINLP).
     \item We introduce an innovative design strategy for the UAV's radar beam, comprising the offline pre-design of its shape and online power allocation for sensing. This approach significantly reduces computational complexity, as the system focuses exclusively on optimizing the scaling factor during the online phase. Particularly, for sensing, our emphasis is on synthesizing a concentrated beam with minimal sidelobes in the offline phase and guaranteeing a required accumulated SNR, facilitated by precise UAV hovering directly above the target, during sensing.
	    \item We develop an alternating optimization (AO) based resource allocation algorithm to solve the formulated non-convex MINLP optimization problem. In particular, we obtain a low-complexity sub-optimal solution  by exploiting semi-definite relaxation, big-M method, and successive convex approximation (SCA). Moreover, we utilize the penalty approach for penalizing the objective function to ensure the equality constraint introduced by the required hovering of the UAV precisely above the target during sensing and for recovering the binary sensing indicator variables, enhancing the efficiency of the solution. 
     
	    \item Our simulation results highlight the benefits of positioning the BS in close proximity to the sensing targets, facilitating efficient data offloading and accurate sensing. Additionally, our results affirm the effectiveness of the proposed algorithm in ensuring precise hovering directly above the targets during sensing, benefiting accurate and reliable target detection.
	\end{itemize}
\textit{Notations:} In this paper, matrices and vectors are denoted by
boldface capital letters $\mathbf{A}$ and lowercase letters $\mathbf{a}$, respectively. $\mathbb{R}^{N\times M}$ and $\mathbb{C}^{N\times M}$ denote the spaces of $N\times M$ real-valued and complex-valued matrices, respectively. $\mathbf{A}^T$,~$\mathbf{A}^H$, $\text{Rank}(\mathbf{A})$, and $\text{Tr}(\mathbf{A})$ are the transpose,~Hermitian, rank, and trace of matrix $\mathbf{A}$, respectively. $\mathbf{A}\succeq\mathbf{0}$ indicates a positive semidefinite matrix. $\mathbf{I}_N$ is the $N$-by-$N$ identity matrix. $|\cdot|$ and $||\cdot||_2$ denote the absolute value of a complex scalar and the $l_2$-norm of a vector, respectively. $\mathbb{E}[\cdot]$ denotes statistical expectation. $\sim$ and $\overset{\Delta }{=}$ stand for ``distributed as'' and ``defined as'', respectively. The distribution of a circularly symmetric complex Gaussian random variable with mean $\mu$ and variance $\sigma^2$ is denoted by $\mathcal{CN}(\mu ,\sigma^2)$. The gradient vector of function $f(\mathbf{x})$ with respect to $\mathbf{x}$ is denoted by $\nabla_{\mathbf{x}} f(\mathbf{x})$.


\begin{figure}
	\centering
	\includegraphics[width=0.85\linewidth]{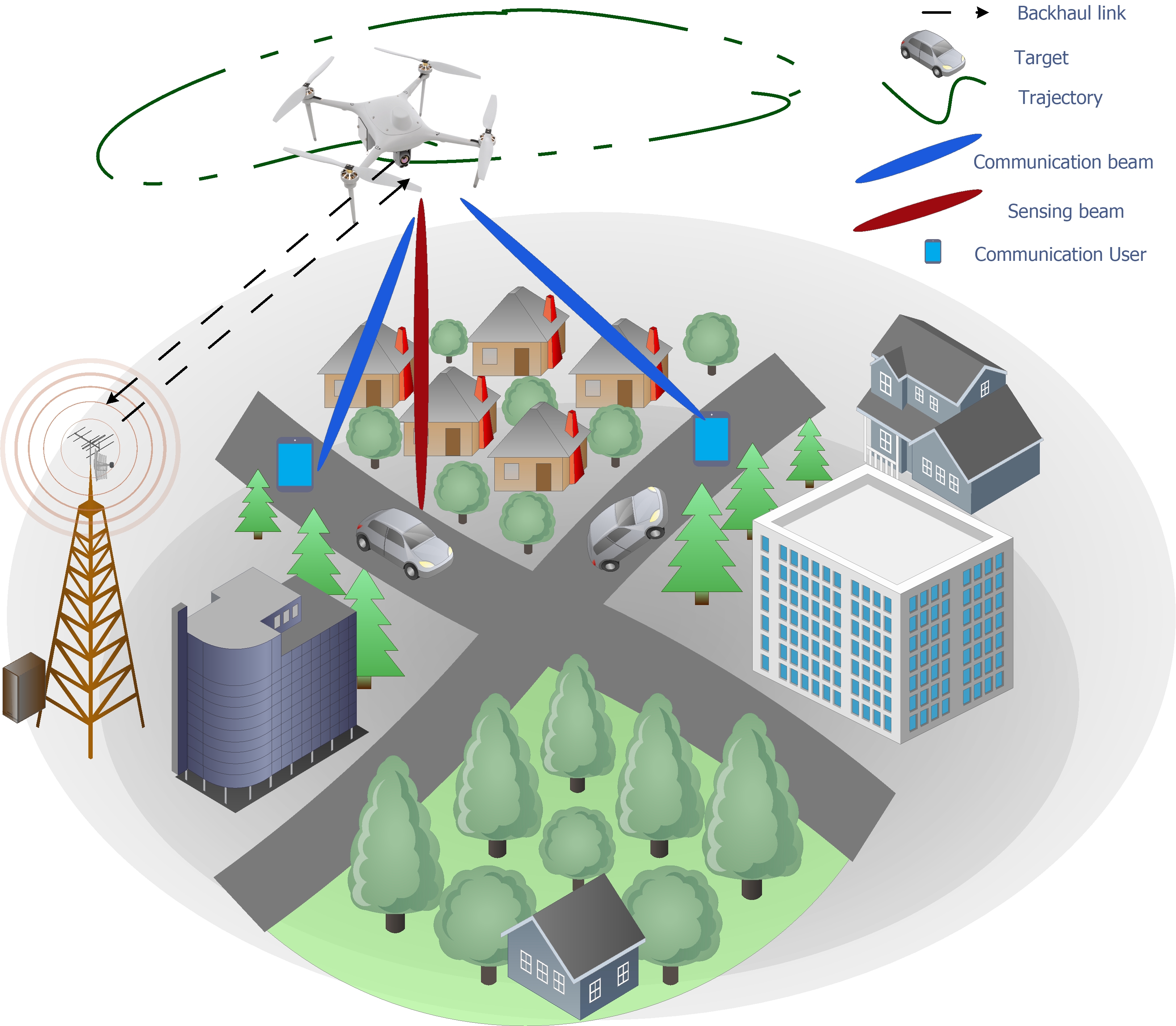}
	\caption{ \small Joint communication and sensing in UAV-assisted network comprising  $E=2$ sensing targets and $K=2$ communication users.   }\label{fig:sys}
	\end{figure}
 \vspace{-5mm}
\section{System Model}
	In this paper, we consider a rotary-wing UAV-assisted ISAC system, which provides downlink communication services for $K$ communication users and senses $E$ potential targets, as depicted in Fig.~1. To cater to the limited computational capabilities of the UAV, ensure low latency data processing, and enable real-time mission monitoring, the UAV offloads the sensing data to a ground BS for further processing \cite{Mag_Qing}. The operation of the system follows a two-step process. The communication data is transmitted from the BS to the UAV, and subsequently, the UAV relays this data to the users. In certain time slots, the UAV switches to the sensing mode, receiving echo signals reflected by the sensing target. These echo signals are first compressed locally at the UAV, reducing the amount of data that needs to be offloaded to the BS. The compressed radar data are then transmitted to the ground BS. Finally, the BS performs central processing for target recognition based on the received compressed sensing data. 
 
 The UAV's total flying time $ T $ is divided into $ N $ time slots of duration $ \delta_t =\frac{T}{N}$. Each time slot is assumed to be sufficiently small, such that the location of the UAV is approximately constant during a time slot to facilitate efficient trajectory and beamforming design for ISAC. 
In the subsequent subsections, we present the proposed ISAC framework in detail. We start by explaining the proposed frame structure. Then, we describe the signal model, including its radar and communication components, before modeling the backhaul links. Finally, we address the power consumption of the UAV, including the power required for local processing, offloading, and flying.
\subsection{ISAC Frame Structure for UAV}
\begin{figure}[t]
	 	\centering
	\includegraphics[width=0.85\linewidth]{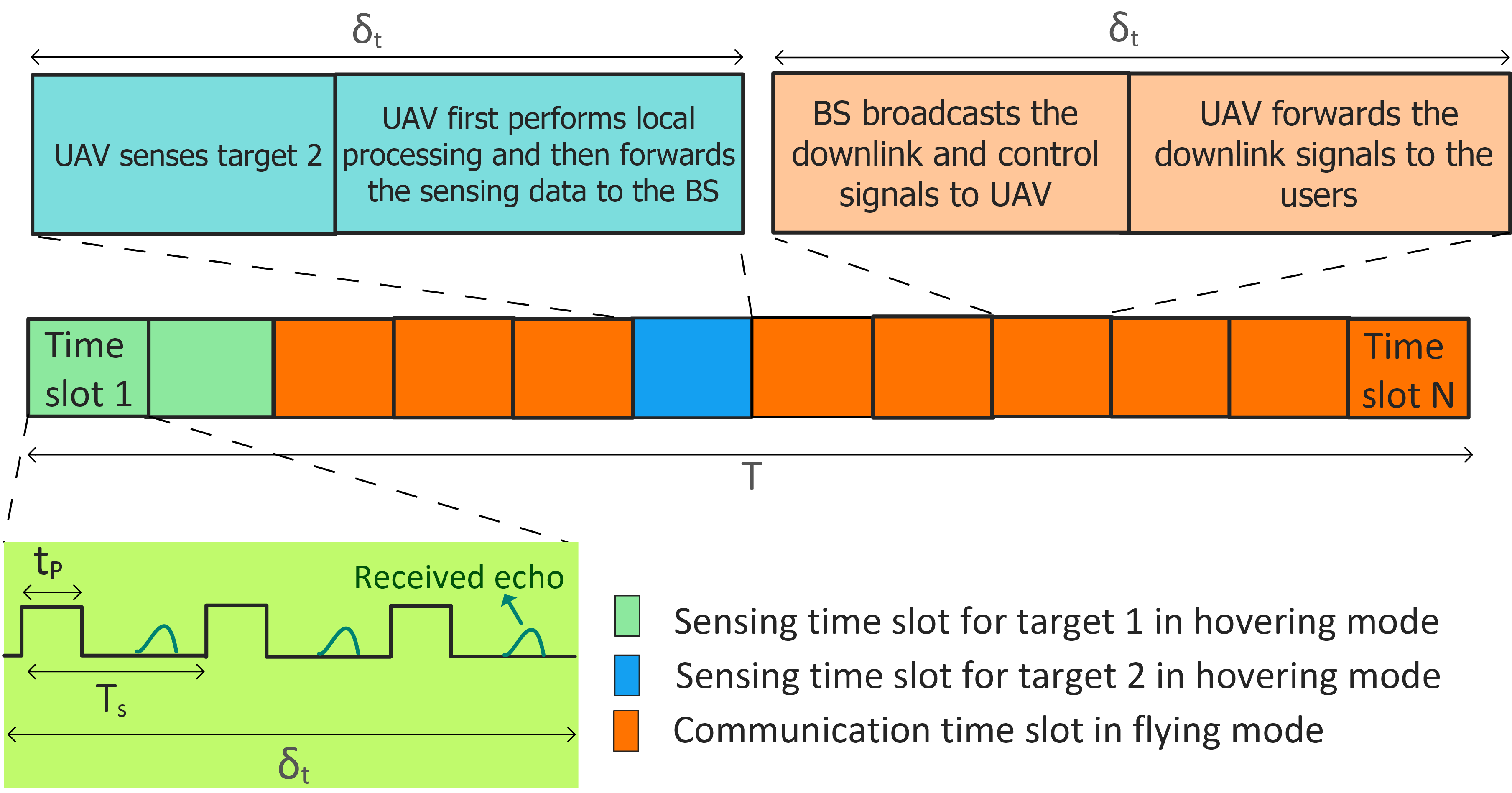}
	 	\caption{ \small Proposed ISAC frame structure where $T$ is the total flying time.}.\label{frame}
   \vspace{-8mm}
	 	\end{figure}

In the proposed UAV-ISAC frame structure, separate and dedicated time slots are employed for sensing and communication, as shown in Fig. \ref{frame}. This strategy minimizes interference, ensuring the UAV's efficient execution of both operations without compromising quality. The UAV communicates with the communication users in the non-sensing time slots, while it senses the target during dedicated sensing time slots, whose number is limited to $N_{s}^{\max}$ to allow for enough time for communication. During sensing, one target is sensed at a time to maximize sensing performance by focusing the beam pattern on the target. The specific time slots in which sensing is performed are determined as part of the optimization process. To this end, we introduce the sensing indicator $\alpha_{e,n}$ for target $e$,  $e \in \{1,...,E\}$. If $\alpha_{e,n}=1$, target $e$ is sensed in the $n$-th time slot; otherwise, $\alpha_{e,n}=0$. Here, we force the UAV to hover above the sensing target. This choice offers several advantages. First, with the UAV hovering above the target, a fixed beam pattern can be designed, eliminating the need for continuous adjustment based on the UAV's flight path, which simplifies the design process. Second, hovering during sensing helps mitigate UAV-induced Doppler shifts, simplifying the signal processing of the sensing data. Third, it minimizes the impact of interference and multi-path effects, resulting in higher sensing performance which allows the UAV to focus the beam pattern with maximum accuracy on the target, enabling the system to extract vital information with optimal efficiency. The assumption of UAV hovering above the target during sensing is justified by its applicability in various real-world scenarios, including vital sign detection through radar technology, where precise UAV positioning is crucial for reliable measurements. These benefits make hovering during sensing preferable in practical UAV-based radar systems\cite{UAVr}.
After sensing, the sensing data is offloaded to the ground BS via a backhaul link to leverage the ground BS's computational capabilities. This offloading reduces latency and enhances sensing precision.
  
 \textit{Remark 1}: The proposed UAV-enabled ISAC system employs separate beams for sensing and communication. This approach avoids the complexity that would arise from a joint beam optimization for sensing and communication, which is particularly challenging for the dynamic conditions associated with UAVs. By assigning separate beams for each task, we can tailor the beam characteristics such as beamwidth, directionality, and allocated power to meet the unique demands of sensing and communication, respectively \cite{ISAC6G}. This not only simplifies the system design but also enhances overall efficiency by reducing the interference between both tasks \cite{niu2024interference}. Separate optimization allows for precise targeting during sensing and broad coverage for communication, avoiding the compromises required in a single-beam design and ensuring better resource utilization. This strategy is particularly effective in mitigating SI, a significant challenge in systems where ongoing communication may impair the received radar echoes\cite{niu2024interference}.
  
  \textit{Remark 2}: The proposed UAV-enabled ISAC system employs single-target sensing as the UAV is required to hover directly above the target during sensing. Compared to multi-target sensing, where multiple targets are sensed concurrently, this approach enhances detection accuracy and reduces interference as a more focused sensing beam can be employed \cite{Regression_beam,Interference_UAV}. Furthermore, for widely-spaced targets, the design of efficient multi-target beams becomes challenging and single-target sensing is preferable.
\subsection{Radar Signal}
The UAV is equipped with a uniform linear array (ULA) with $M$ antennas for communication and sensing. We adopt a three-dimensional (3D) Cartesian coordinate system where the horizontal location of the UAV in time slot $n$ and the location of the potential target on the ground are denoted by ${\mathbf{q}}[n] = {\big[q_x[n],q_y[n]\big]^{T}}$ and ${\mathbf{d}}_{e} = {\big[d_{x_e}, d_{y_e}\big]^{T}} \in {\mathbb{R}^{2 \times 1}}$, respectively. The value of ${\mathbf{d}}_{e}$, $e \in \{1,...,E\}$,  is predetermined based on the specific sensing tasks\footnote{The value of ${\mathbf{d}}_{e}$ could be set based on an estimated location for target tracking or it could be a fixed location in the region of interest for target detection \cite{UAVISAC1,ThUAVISAC}.}. While having initial knowledge of the target's location is valuable, the complete sensing data acquisition involves a two-step process. Initially, the UAV acquires rough information to initiate the sensing process. Subsequently, it performs detailed sensing to gather sufficient data on the target which is then offloaded to the BS for further processing. In the considered system, the UAV's initial knowledge guides the sensing process, ensuring efficient and accurate data acquisition. The offloading to the BS facilitates in-depth analysis and contributes to a comprehensive understanding of the target including the target’s location, angle, shape, velocity, size, and parameters. Moreover, it is assumed that the UAV flies in the $ x-y $ plane at fixed altitude $ H $. While hovering over a given target, the UAV emits a narrow beam toward the direction of the target to extract information from the target. The radar signal $\mathbf{s}_{0} \in \mathbb{C}^{M \times 1}$ with covariance matrix $\mathbf{R}[n] = \mathbb{E}[\mathbf{s}_0[n]\mathbf{s}_0^H[n]] \succeq \mathbf{0}$ is transmitted towards the given target in a sensing slot.~The transmit beam pattern gain from the UAV in the direction of target $e$ is given by 
 \begin{align}\mathcal{P}(\mathbf{R},\mathbf{q}[n],\mathbf{d}_{e})=\mathbf{a}^H(\mathbf{q}[n],\mathbf{d}_{e})~\mathbf{R}[n]~\mathbf{a}(\mathbf{q}[n],\mathbf{d}_{e}),
 \end{align}where 
\begin{align} \label{equ:steering}
	\scalemath{0.9}{\mathbf{a}(\mathbf{q}[n],\mathbf{d}_{e}) = \big[1,e^{j 2\pi \frac{\hat{d}}{\lambda} \cos (\theta(\mathbf{q}[n],\mathbf{d}_{e}))},...,e^{j 2\pi \frac{\hat{d}}{\lambda} (M-1) \cos(\theta(\mathbf{q}[n],\mathbf{d}_{e}))}\big]^T}
\end{align} 
is the steering vector of the ULA equipped at the UAV, $\theta(\mathbf{q}[n],\mathbf{d}_{e})=\arccos\big(\frac{H}{\sqrt{\|\mathbf{q}[n]-\mathbf{d}_{e}\|^{2}+H^{2}}}\big)$ is the angle of departure corresponding to target $e$, $\lambda$ is the carrier wavelength, and $\hat{d}$ denotes the spacing between two adjacent UAV antennas.

We adopt a two-phase strategy to optimize the design of the beam pattern and to maximize the quality of sensing. 

\subsubsection{Shape of the Sensing Beam} In the offline phase, a highly directional sensing beam pattern is designed, efficiently catering to specific constraints required for optimal sensing. To facilitate high-quality sensing during the UAV's hovering phase, the desired sensing location has to be illuminated by an energy-focused beam with low side lobe leakage, facilitating the separation of the desired echoes and clutter. The UAV employs a pre-designed highly directional sensing beam pattern characterized by a specific covariance matrix that defines the desired waveform. To this end, we discretize the angular domain $[-\frac{\pi}{2}, \frac{\pi}{2}]$ into $L$ directions and generate the ideal beam pattern $\{\mathcal{D}{(\theta_l)}\}_{l=1}^{L}$, where  $\mathcal{D}(\theta_l)$ denotes the beam pattern power in direction $l$, which is given by
 \[
\mathcal{D}{(\theta_l)}=
\begin{cases}\label{desired_a}
1, &\theta_{e}-\Delta\leq \theta_l \leq\theta_{e}+\Delta, \:\:\\
  0, & \text{otherwise},
\end{cases}
\]
where $2\Delta$ is the beamwidth used to sense one target\footnote{~In practice, the beamwidth $2\Delta$ has to be carefully chosen as a larger $\Delta$, on the one hand, increases coverage and tolerance with respect to uncertainties regarding the target position, while on the other hand, it also compromises the radar's sensitivity and accuracy by increasing the amount clutter captured alongside the desired signals\cite{mu-mimo-jsc,Probing}.} \cite{Probing}, and $\theta_{e}$ is the angle of departure corresponding to target $e$. In the hovering state, the horizontal distance between the UAV and the target is zero, resulting in an angle of zero degrees between the UAV and the target, i.e., $\theta_{e}=0$. Consequently,  
to shape the beam, we adopt the minimum square error (MSE) criterion, which is given by \cite{Probing}
\begin{align}\label{Rd}
\underset{\rho_{0},\mathbf{R}_{d}}{\text{minimize}}~  &\frac{1}{L}\sum_{l=1}^L \bigg|\rho_{0}\mathcal{D}(\theta_l) -\mathbf{a}^H(\theta_l) \mathbf{R}_{d}\mathbf{a}(\theta_l)\bigg|^2\\
     \text{s.t.}~&\text{Tr} (\mathbf{R}_{d})=1\nonumber,\\
     & \mathbf{R}_{d}=\mathbf{R}_{d}^{H},~\mathbf{R}_{d}\succeq \mathbf{0}\nonumber, 
 \end{align}
 where $\rho_{0}$ is a scaling factor. Problem \eqref{Rd} is a semi-definite quadratic programming problem and can be efficiently solved in polynomial time by CVX.
 \subsubsection{Scaling Power of the Sensing Beam} In the online phase, we scale and configure the beam pattern in real time. The employed radar beam pattern, denoted as $\mathbf{R}_{d}$, is obtained from \eqref{Rd}, and remains fixed and does not depend on time slot $n$. As mentioned before, the beam pattern is specifically designed for scenarios when the UAV hovers directly above the target. We introduce scaling factor $p^{\text{Rad}}[n]$, which is applied to the desired radar beam pattern matrix $\mathbf{R}_{d}$ yielding $\mathbf{R}[n]=p^{\text{Rad}}[n]\mathbf{R}_{d}$ for the covariance matrix used for sensing. This scaling factor allows for dynamic adjustment of the beam power during sensing. 

We adopt pulse radar for sensing to ensure reliable echo
detection at the transmitter and to provide flexibility in adjusting the sensing range. According to pulse radar theory, the sensing range is contingent upon the duration of the sensing pulse and
the time taken to listen for the received echo \cite{Skolnik2008}. Consequently, the system designer meticulously divides the available sensing time into two components to ensure dependable echo detection at the transmitter. As a result, each sensing slot comprises multiple scan rounds, within each of which the UAV transmits a scanning pulse lasting for a duration of $t_{p}$, as shown in Fig 2. Following this transmission, the UAV switches to the listening mode to receive the target's echo corresponding to the transmitted pulse. Consequently, each sensing round operates at a specific pulse repetition frequency (PRF). In particular, $ T_{\text{s}} = t_p+t_o $ represents the duration of each sensing round, where $t_{o}$ corresponds to the duration of the listening mode (reception duration of the received radar echo). The number of sensing rounds per sensing time slot is then given by $N_s=\frac{\delta_t}{T_{\text{s}}}$. Furthermore, we assume that the same sensing signal $\mathbf{s}_{0}[n]$ is used in each sensing round within a given time slot and that the channel remains constant in each sensing round in a given time slot. The latter assumption is justified as the UAV hovers and the considered targets are stationary. In each sensing round, the radar transmits a pulse of duration $t_{p}$. As a result, the echo signal received from target $e$ at the UAV in sensing round $n_{s}$, $1\leq n_{s}\leq N_{s}$, of time slot $n$ is given by
\begin{align}
\mathbf{r}_{e}[n,n_{s}]&= \sqrt{\frac{t_p}{\delta_t}} \mathbf{H}_{e}[n]
\mathbf{s}_{0}[n] + \mathbf{z}[n,n_{s}],
\end{align}
where $\mathbf{z}[n,n_{s}]\sim\mathcal{C}\mathcal{N}(\mathbf{0},\sigma^{2}_{e}\mathbf{I}_{M})$ is the received additive white Gaussian noise (AWGN) at the UAV in sensing round $n_{s}$ of time slot $n$. Furthermore, $\mathbf{H}_{\mathrm{e}}[n]$ is the round-trip channel matrix in time slot $n$, which is given by
\begin{align}
\mathbf{H}_{e}[\hspace{-0.25mm}n\hspace{-0.25mm}] \hspace{-1mm}=\hspace{-1mm}\frac{\epsilon_{e}\hspace{-0.25mm}[\hspace{-0.25mm}n\hspace{-0.25mm}]\beta_0}{2\Psi_{e}[n]} \mathbf{a}{(\mathbf{q}[n],{\mathbf{d}}_e)}\mathbf{a}^{H}{(\mathbf{q}[n],{\mathbf{d}}_e)},
\end{align}
where $\beta_{0}$ denotes the channel power gain at the reference distance of $d_{0} = 1$~m and $\Psi_{\mathrm{e}}[n]={{{\sqrt{\left\| {{\mathbf{q}[n]} - {\mathbf{d}_e}} \right\|^2+H^2}}}}$. Moreover, $\epsilon_{e}\hspace{-0.25mm}[\hspace{-0.25mm}n\hspace{-0.25mm}]\hspace{-1mm} =\hspace{-1mm} \sqrt{\hspace{-1mm}\frac{\vartheta_{e}}{4\pi \Psi^2_e[n]}}$ denotes the reflection coefficient of target $e$, and $\vartheta_{e}$ is the radar cross-section of target $e$ \cite{Radar}. Fraction $\frac{t_{p}}{\delta_{t}}$ represents the proportion of time each pulse occupies within a time slot. To combine the signals received at the different antennas, the UAV applies receive beamforming vector \(\mathbf{u}[n]= \frac{\mathbf{a}(\mathbf{q}[n], \mathbf{d}_e)}{\|\mathbf{a}(\mathbf{q}[n], \mathbf{d}_e)\|_{2}}\). Consequently, the combined echo signal at the UAV is given by $
y_{e}[n,n_{s}] = \mathbf{u}^{H}[n] \mathbf{r}_{e}[n,n_{s}]$. Signal \( y_{e}[n,n_{s}] \) is then forwarded to the ground BS. At the BS, the signals received in all the sensing rounds of a single time slot are coherently combined to enhance the SNR. The resulting combined signal in time slot \(n\) is given by 
\begin{align}
y_{e}[n] = \sum_{n_{s}=1}^{N_{s}} y_{e}[n, n_{s}]=&\underbrace{N_{s}\sqrt{\frac{t_p}{\delta_t}}\mathbf{u}^{H}[n] \mathbf{H}_{e}[n]  \mathbf{s}_{0}[n]}_{\triangleq r[n]} + \nonumber\\&\underbrace{\sum_{n_{s}=1}^{N_{s}}\mathbf{u}^{H}[n] \mathbf{z}[n, n_{s}]}_{\triangleq z_{\text{eff}}[n]}. 
\end{align}
Consequently, in time slot $n$, the radar output SNR for detection of target $e$ at the BS is given by
\begin{align}
\gamma_{e}[n] = \frac{N_s \frac{t_p}{\delta_t}  \mathbf{u}^{H}[n]\mathbf{H}_{e}[n] \mathbf{R}[n] \mathbf{H}_{e}^{H}[n]\mathbf{u}[n]}{\sigma^{2}_{e}\mathbf{u}^{H}[n]\mathbf{u}[n]},
\end{align} 
which can be further simplified to 
\begin{align}
\gamma_{e}[n]&=\frac{\vartheta_{e}\beta_{0}^{2}N_s\frac{t_p}{\delta_t}\mathbf{a}^{H}{(\mathbf{q}[n],{\mathbf{d}}_e)}
\mathbf{R}[n]{\mathbf{a}{(\mathbf{q}[n],{\mathbf{d}}_e)}}}
{16\pi \Psi^4_{e}[n]\sigma^{2}_{e}}.
\end{align}
To further improve the SNR, accumulation across multiple time slots is beneficial to average out variations in channel quality, interference, and noise characteristics \cite{UAV_SNR, Richards2010}. The combined signal for detection of target $e$ can be represented as  $y_{e,\text{total}} = \sum_{n=1}^{N} b_{n} y_{e}[n]=\sum_{n=1}^{N} b_{n}r[n]+\sum_{n=1}^{N} b_{n}z_{\text{eff}}[n]$, where $b_{n}$ is the combining weight. Here, $r[n]$ represents the signal obtained in time slot $n$ after the accumulation of all echoes, while $z_{\text{eff}}[n]$ denotes the resulting effective noise. In order to maximize the SNR across the $N$ time slots, while taking into account that sensing is performed only in the dedicated sensing time slots, we choose the combining weights as 
   $ b_{n}=\alpha_{e,n}r^{*}[n]$. This can be interpreted as maximum ratio combining (MRC) with additional selection. For given $\mathbf{s}_{0}[n]$, the resulting accumulated sensing SNR can be obtained as $\widetilde{\Gamma_{e}}=\frac{1}{N_{s}\sigma^{2}_{e}}\sum_{n=1}^{N}\alpha_{e,n}|r[n]|^{2}$. Then, after taking the expectation with respect to $\mathbf{s}_{0}[n]$, we obtain for the accumulated sensing SNR
\begin{align}\label{Gammae}
 \Gamma_{e}=\sum_{n=1}^{N}\alpha_{e,n} \frac{\vartheta_{e}\beta_{0}^{2}{N_s\frac{t_p}{\delta_t}}\mathbf{a}^{H}(\mathbf{q}[n],\mathbf{d}_{e}) \mathbf{R}[n]\mathbf{a}(\mathbf{q}[n],\mathbf{d}_{e})}{16\pi \Psi^4_{e}[n]\sigma^{2}_{e}}.
\end{align}
To achieve satisfactory sensing performance\footnote{In the considered ISAC system, the UAV focuses solely on collecting and transmitting high-quality sensing data, while detection is handled by the ground BS. Increasing the accumulated sensing SNR is beneficial for both the probability of detection and the probability of false alarm. Therefore, in this paper, we aim to guarantee a minimum accumulated sensing SNR. Depending on the sensing objective, the BS can set a proper decision threshold to adjust the tradeoff between the probability of detection and the probability of false alarm.}, we require the accumulated sensing SNR of target $e$ to be higher than a pre-defined minimum threshold, denoted by $\text{SNR}^{\text{th}}_{e}$,~i.e., 
\begin{align}\label{SNR_C}
\Gamma_{e}\geq\text{SNR}_{e}^{\text{th}}.
\end{align}

\textit{Remark 3:} The proposed UAV-based ISAC system employs pulse radar due to its superior echo detection capabilities compared to continuous wave (CW) radar\cite{Radar,Levanon2004}. Pulse radar systems excel in measuring distances accurately by calculating the time delay between pulse emission and echo reception, ensuring precise target detection \cite{Radar}. Additionally, by adjusting pulse duration and pulse repetition frequency the peak power can be enhanced which is beneficial for detecting distant or low-reflectivity targets \cite{Richards2010, Levanon2004}. By emitting signals intermittently, pulse radar systems avoid the continuous background noise and interference from other systems, achieving high sensitivity for detection \cite{Richards2010, Levanon2004}. In comparison, CW radar systems continuously transmit signals, which can lead to challenges in distinguishing between transmitted and received signals due to SI. CW radar is typically used for measuring velocity rather than distance, as it does not provide direct range information without additional modulation\cite{CW0,CW2}. Nevertheless, extending the proposed UAV-based sensing framework to CW radar and velocity estimation is an interesting topic for future work. 

\textit{Remark 4:} The proposed ISAC system utilizes the accumulated sensing SNR in \eqref{Gammae} as performance metric for sensing. Accumulating the SNR over multiple sensing cycles provides robustness against transient noise spikes and interference, ensuring more reliable detection \cite{UAV_SNR, Richards2010}. The use of MRC for combining samples $y_{e}[n]$ maximizes the accumulated SNR and serves as an upper bound for other combining schemes. If a different combining scheme is used, \eqref{Gammae} and \eqref{SNR_C} may still be used but the predefined threshold SNR$^{\text{th}}_{e}$ has to be increased to account for the less efficient combining.  Alternatively, the accumulated sensing SNR for the employed suboptimal combining scheme could be derived and used instead of $\Gamma_{e}$ in \eqref{SNR_C}.

\subsection{Communication Signal}
The horizontal location of the $K$ communication users is denoted by ${\mathbf{d}}_{k}= {\big[d_{x_k}, d_{y_k}\big]^{T}}$. Consequently, the channel vector between the UAV and user $k$ is denoted by $\mathbf{h}_{k}$, and given by
 \begin{align}
	{\mathbf{h}_{k}}{[n]} = \frac{{{\beta _0\mathbf{a}{(\mathbf{q}[n],{\mathbf{d}}_k[n])}}}}{{{{\sqrt{\left\| {{\mathbf{q}[n]} - {\mathbf{d}_k}} \right\|^2+H^2}}}}},
 \end{align}
 based on the free space channel model. In the non-sensing slots, the UAV transmits simultaneously information symbols $c_k[n]$, $c_{k}\sim \mathcal{CN}(0,1)$, $k \in \{1,...,K\}$, to the $K$ communication users. Then, the received signal at user $k$ can be written as
\begin{equation}
	     y_{k}[n]=\mathbf{h}^{H}_k[n]\sum_{j=1}^K \mathbf{w}_j[n] c_j[n]+z_{k}[n],
      \end{equation}
where $\mathbf{w}_j[n] \in \mathbb{C}^{M \times 1}$ denotes the transmit beamforming vector and $z_{k}\sim\mathcal{CN}(0,\sigma^{2}_{k})$ is the AWGN at user $k$. As a result, the received SINR at user $k$ in time slot $n$ is given by  
		\begin{equation}\label{sinr}
		{\gamma_{k}}{[n]} = \dfrac{\big|\mathbf{h}_k^{H}[n]\mathbf{w}_k[n]\big|^2}{\sum_{i\neq k}^{}\big|\mathbf{h}_k^{H}[n]\mathbf{w}_{i}[n]\big|^2+\sigma^2_k}.
		\end{equation}


\subsection{Backhaul Model}
\subsubsection{Radar pulse}
Based on the PRF, the minimum and maximum sensing ranges, for which the UAV can detect a target, are given by\cite{Radar}
\begin{align}
{R}_{\min}= \frac{ct_p}{2}~~\text{and}\:\:\:{R}_{\max}= \frac{ct_o}{2}.
\end{align}
\subsubsection{UAV-BS}
 In each sensing time slot, the UAV has to first sample and quantize the received echo signal based on the desired sensing resolution, and then forward the quantized data to the BS. Consequently, we model the backhaul capacity required for conveying the sampled and quantized echoes from the UAV to the BS during each time slot, as follows \cite{Backhaul}
 
 \begin{align} \label{rate_production}
R_{\text{Pr}}= \frac{N_{s}N_{b}(R_{\max}-R_{\min})}{\Delta{R}~\delta_{t}~W_f}, 
\end{align}
where $N_{b}$ is the number of bits needed to characterize a quantized value of the echo signal, $\Delta{R}$ is the resolution of the radar in meters determined by the pulse width, type of target, and efficiency of the radar\cite{Radar}, and $W_{f}$ is the bandwidth of the backhaul link. By neglecting the effect of quantization errors, we simplify the analysis and assume that the quantization process introduces negligible distortion.After compression of the radar data at the UAV, the compressed data are offloaded to the BS for further processing and analysis. We model the achievable data rate between the UAV and the BS based on an equivalent single-input single-output (SISO) link as follows\footnote{Since the channel between the UAV and the BS is LoS, only one spatial degree of freedom is
available. Accordingly, the backhaul channel matrix is rank-one with a unique non-zero singular value.}
\begin{align}
R_{\text{U-B}}[n]=\log_2\Big(1+\frac{\alpha_{e,n}p^{\text{Off}}[n]\lambda_{1}^{2}[n]}{\sigma^{2}_{B}}\Big),
\end{align}
where $\lambda_{1}[n]=\frac{\sqrt{\beta _0 G_{T}}}{{{{{\sqrt{\left\| {{\mathbf{q}[n]} - {\mathbf{q}_b}} \right\|^2+H_b^2}}}}}}$, and $G_{T}$ is the antenna gain for the backhaul link. Also, $H_b=H-H_{\text{BS}}$, where $H_{\text{BS}}$ is the height of the BS.~$p^{\text{Off}}[n]$ is the transmission power needed for offloading the radar data from the UAV to the BS and $\sigma^{2}_{B}$ is the variance of the noise at the BS. To guarantee successful real-time communication between the UAV and the BS, the production rate (the rate at which sampled and quantized echoes are conveyed from the UAV to the BS) should be smaller than the achievable rate of the backhaul link. Thus, the following inequality must hold in the sensing  slots 
\begin{align}
\text{C}4: R_{\text{U-B}}[n]\geq \alpha_{e,n}\iota R_{\text{Pr}},
\end{align}
where $\iota$, $0<\iota<1$, denotes the data compression factor resulting from the local compression carried out by the UAV. 
\subsubsection{BS-UAV} Besides, the backhaul constraint required for offloading of sensing data, the link between the BS to the UAV must also satisfy a minimum QoS requirement to ensure successful data transmission to the users
\begin{align}
&\text{C}5:R_{\text{B-U}}[n]=\log_2\Big(1+\frac{p_{\text{BS}}[n]\lambda^{2}_{1}[n]}{\sigma^{2}_{}}\Big)\geq \nonumber\\&\sum_{k=1}^{  K}R^{k}_{\min}\Big(1-\sum_{e=1}^{E}\alpha_{e,n}\Big),
\end{align}
where $R^{k}_{\min}$ is a minimum QoS requirement for communication user $k$, and $\sigma^{2}$ is the AWGN variance at the UAV.


\subsection{Power Consumption Model}
Besides, the power consumption incurred for data transmission and sensing, the UAV also consumes power for offloading, local data processing, and flying.  
\subsubsection{Offloading Power Consumption of the UAV}
In the context of our system, offloading is a critical strategy to manage the significant amount of radar data generated during the sensing phase. The UAV needs to efficiently transmit this data to the BS with transmit power $p^{\text{Off}}[n]$ for further processing.

\subsubsection{Local Power Consumption of the UAV}
We model the power consumption required for local data processing, i.e., for data compression at the UAV, as follows  \cite{Eloc}
\begin{align}
\mathcal{P}_{\text{Loc}}=a~f^{3}_{\text{Loc}}, 
\end{align}
where $a$ is a constant related to the hardware architecture of the UAV and $f_{\text{Loc}}$ (cycles/sec) is the local computation resource of the UAV.
	\subsubsection{Aerodynamic Power Consumption of the UAV}The propulsion power consumption depends on the flying mode of the UAV \cite{DUAV,Rotary}. In particular, the aerodynamic power consumption for rotary-wing UAVs is a function of the flight velocity $\mathbf{v}[n]\in \mathcal{R}^{2\times 1}$\cite{Rotary}.  The total power consumption in time slot $n$ can be written as
 \begin{align}
P_{\text{aero}}(\mathbf{v}[n]) &= \sum_{e=1}^{E}\alpha_{e,n}P_{\text{hover}}[n]+\Big(1-\sum_{e=1}^{E}\alpha_{e,n}\Big)P_{\text{fly}}(\mathbf{v}[n]),\label{eqn:P_flight_I}     
 \end{align}
where $P_{\text{hover}}=P_o+P_i$ and $P_{\text{fly}}$=$
P_o \bigg(\frac{3 \|\mathbf{v}[n]\|^2 }{\Omega^2 r^2} \bigg) + P_i  \bigg[\left( \sqrt{1+\frac{\|\mathbf{v}[n]\|^4}{4 v_0^4}}-\frac{\|\mathbf{v}[n]\|^2}{2v_0^2}\right)^{1/2}-1\bigg]+ \frac{1}{2} r_0 \rho s A_{\mathrm r} \|\mathbf{v}[n]\|^3.$
The parameters of the power consumption model are defined in Table \ref{notations}\cite{Rotary}.
 \begin{table}[t]
\centering
\scriptsize
\caption{Parameters in the power consumption model\cite{Rotary}.} \label{notations} 
\begin{tabular}{ c | c }
  \hline			
  Notations & Definitions \\ \hline
  $\Omega=300$ & Blade angular velocity in radians/second \\
  $r=0.4$ & Rotor radius in meter \\
  $\rho=1.225$ & Air density in $\mathrm{kg/m^3}$ \\
  $s=0.05$ & Rotor solidity in $\mathrm{m^3}$ \\
  $A_{\mathrm r}=0.503$ & Rotor disc area in $\mathrm{m^2}$ \\
  $P_o=80$ & Blade profile power during hovering in Watt\\
  $P_i=88.6$ & Induced power during hovering in Watt\\
  $v_0=4.03$ & Mean rotor induced velocity in forward flight in m/s\\
  $r_0=0.6$ & Fuselage drag ratio \\
  $P_{\text{static}}=0.3$~W & Circuit power consumption
of RF chain\\
\hline
\end{tabular}
\vspace*{-6mm}
\end{table}
	\section{Problem Formulation}
	In this paper, we aim to minimize the average power consumption of the UAV which includes the transmission power, aerodynamic power consumption, and power consumption for offloading by jointly optimizing the beamforming for communication, the power for sensing ($p^{\text{Rad}}[n]$), the UAV's trajectory ($ \mathbf{q} $), the velocity ($ \mathbf{v} $), and the sensing indicator while guaranteeing the QoS of the communication users as well as the sensing targets.~As a result, the optimization problem is mathematically formulated as follows:
	\begin{align}
	 \mathcal{P}_{1}:&\mathop {{\rm{min}}} \limits_{\scriptstyle{\boldsymbol{\Xi}}}\mathcal{O}bj\triangleq\frac{1}{N}\sum_{n=1}^{N}\bigg(\eta(\sum_{k=1}^{K} \|\mathbf{w}_k[n]\|^2+N_s\frac{t_p}{\delta_{t}}\text{Tr}(\mathbf{R}[n]))+\nonumber\\&\scalemath{0.9}{P_{\text{aero}}(\mathbf{v}[n])+M~P_{\text{static}}+\sum_{e=1}^{E}\alpha_{e,n}\mathcal{P}_{\text{Loc}}[n]+\sum_{e=1}^{E}\alpha_{e,n}p^{\text{Off}}[n]\bigg)}\nonumber\\
	\text{s.t.}~~
& \text{C}1:\Big(1-\sum_{e=1}^{E}\alpha_{e,n}\Big)\sum_{k=1}^{K} \|\mathbf{w}_k[n]\|^2+\nonumber\\&\sum_{e=1}^{E}\alpha_{e,n}N_s\frac{t_p}{\delta_t}\text{Tr}(\mathbf{R}[n]) +\sum_{e=1}^{E}\alpha_{e,n}p^{\text{Off}}[n]\leq P_{\max},\forall n,\nonumber\\
	& \text{C}2:\frac{1}{N}\sum_{n=1}^{N}\Big(1-\sum_{e=1}^{E}\alpha_{e,n}\Big)\log_2(1+\gamma_{k}[n])\geq R_{\min}^{k}, \forall k,\nonumber\\
 &\text{C}3:\sum_{n=1}^{N}\alpha_{e,n}\frac{\vartheta_{e}\beta_{0}^{2}\mathbf{a}^{H}(\theta_{e}){N_s\frac{t_p}{\delta_t}}\mathbf{R}[n]\mathbf{a}(\theta_{e})}{16\pi \Psi^4_{e}[n]\sigma^{2}_{e}}\geq \text{SNR}_{e}^{\text{th}},\forall e,\nonumber\\
 &\text{C}4: R_{\text{U-B}}[n]\geq \alpha_{e,n}\iota R_{\text{Pr}},~\forall e,n,\nonumber\\&\text{C5}:R_{\text{B-U}}[n]\geq \sum_{k=1}^{  K}R^{k}_{\min}\Big(1-\sum_{e=1}^{E}\alpha_{e,n}\Big),~\forall n,\nonumber
 \end{align}
 \begin{align}\label{20}
	&\text{C}6:\sum_{e=1}^{E}\alpha_{e,n}\leq 1, \forall n,~\text{C}7:\sum_{n=1}^{N}\alpha_{e,n}\leq N_{s}^{\max}, \forall e,\nonumber\\
			& \text{C}8:\mathbf{q}[n+1]-\mathbf{q}[n]=\Big(1-\sum_{e=1}^{E}\alpha_{e,n}\Big) \mathbf{v}[n] \delta_t,\forall n,\nonumber\\
	&\text{C}9:\big\|\mathbf{v}[n+1]-\mathbf{v}[n]\big\|\leq a_{\max}\delta_t,\forall n,\nonumber\\
& \text{C}10:\big \|\mathbf{v}[n]\big \| \leq \Big(1
	-\sum_{e=1}^{E}\alpha_{e,n}\Big) v_{\max},\forall n,\nonumber\\
&\text{C}11:\alpha_{e,n} \in \{0,1\}, \forall e,n,~\text{C}12: \alpha_{e,n}\big\|\mathbf{q}[n]-\mathbf{d}_{e}\big\|^{2}=0,
\end{align} 
where $\eta>1$ and $P_{\text{static}}$ denote the power amplifier efficiency
and the circuit power consumption of the radio frequency (RF)
chain of one antenna element, respectively. In optimization problem $\mathcal{P}_{1}$, $\boldsymbol{\Xi}\triangleq\{\mathbf{w}_{k}[n],p^{\text{Rad}}[n],p^{\text{Off}}[n], \alpha_{e,n}, \mathbf{q}[n],\mathbf{v}[n]\}$ is the set of optimization variables. \text{C}1 limits the transmit power of the UAV, where $P_{\max}$ is the maximum transmit power. \text{C}2 guarantees that the average achievable data rate of the communication users in non-sensing slots does not fall below $R^{k}_{\min}$. \text{C}3 ensures that the accumulated sensing SNR at the UAV exceeds a specified minimum threshold, denoted by $\text{SNR}^{\text{th}}_{e}$, necessary for effective target sensing. 
\text{C}4 indicates that the rate of production must not exceed the achievable rate of the backhaul link to ensure real-time communication between UAV and BS. \text{C}5 guarantees that the communication link between the BS and the UAV satisfies the minimum QoS that the UAV has to provide to the users. \text{C}6 ensures that at most one target is sensed in each time slot. \text{C}7 limits the maximum number of time slots used for sensing of each target to $N_{s}^{\text{max}}$. \text{C}8 models the evolution of the trajectory of the UAV based on its flight velocity. Furthermore, 
	 \text{C}9 and \text{C}10 limit the maximum acceleration and velocity of the UAV to $a_{\text{max}} $ and $ v_{\text{max}} $, respectively. \text{C}11 specifies that the sensing indicator is an integer variable. Finally, controlled by the sensing indicator $\alpha_{e,n}$, constraint C12 ensures precise UAV positioning during sensing. When $\alpha_{e,n} = 1$, C12 guarantees that the UAV hovers directly above the target, eliminating any horizontal distance between the UAV and the target. Conversely, when $\alpha_{e,n} = 0$, the constraint does not apply, allowing the UAV to maneuver freely for communications with the ground users.

  \textit{Remark 5:} In the objective function of $\mathcal{P}_{1}$, sensing indicator $\alpha_{e,n}$ is not explicitly included in the communication and sensing transmit power. This is not needed as for the optimal solution of $\mathcal{P}_{1}$, if $\alpha_{e,n}=1$, no transmission power is allocated for communication in time slot $n$, i.e., $||\mathbf{w}_{k}[n]||=0$, as this time slot does not contribute to meeting C2. Similarly, if $\alpha_{e,n}=0$, no transmit power is allocated to sensing i.e.,  $||\mathbf{R}[n]||=0$, as in this case, time slot $n$ does not contribute to meeting C3.

  \textit{Remark 6}: In the proposed UAV-based ISAC system, enforcing the UAV to hover directly above the target during sensing simplifies system design. In particular, the need for beam pattern adjustments during sensing, which would be necessary if the UAV was moving, is avoided\cite{Regression_beam}. In addition, by maintaining a fixed position directly above the target, interference and multipath effects are minimized\cite{Interference_UAV}.
  
\section{Solution of the optimization problem}
Optimization problem $\mathcal{P}_{1}$ is challenging to solve. The challenge primarily stems from the intricate interplay of various system parameters and the non-convexity introduced by constraints $\text{C1}-$$\text{C5}$,~$\text{C11}$, $\text{C12}$, and the UAV's power consumption model in the objective function.  Moreover, the inclusion of binary sensing indicator $\alpha_{e,n}$ in C1-C5, C8, C11, and C12 transforms the optimization problem into an MINLP problem, further enhancing its complexity. Additionally, satisfying the equality constraint C12 poses a significant challenge in efficiently solving the formulated problem. The presence of UAV trajectory variables in the exponential functions in the steering vectors aggravates the difficulty of the joint UAV trajectory and beamforming optimization problem. Consequently, finding a globally optimal solution in polynomial time for this problem is a formidable task. The problem involves multiple variables and constraints, making it inherently complex. By decomposing it into two sub-problems, we break down the complexity, promoting a manageable and efficient solution. Thus, to strike a balance between computational complexity and performance, we propose a suboptimal solution approach using an iterative algorithm based on the AO technique. Beamforming for the communication users, the power of the sensing beam, the power for offloading along with binary sensing indicators are primarily related to communication and sensing aspects. These variables can be optimized relatively independently of the UAV's trajectory and velocity. Separating these aspects into a sub-problem enables parallelization, potentially speeding up the overall optimization process. Therefore, in a first step, we jointly optimize the communication and sensing variables, using a combination of semi-definite programming (SDP), Big-M method, SCA, and the penalty approach. Then, in a second step, we jointly optimize the UAV trajectory and velocity, utilizing SCA, Big-M, and the penalty approaches. Despite being suboptimal, this approach simplifies the problem and allows us to significantly reduce power consumption while meeting the prescribed communication and sensing performance requirements. The key steps for finding the solution to the considered overall optimization problem $\mathcal{P}_{1}$ are illustrated in Figure 3.

\begin{figure}
	\centering
\hspace{0.25cm}\includegraphics[width=1.5\linewidth]{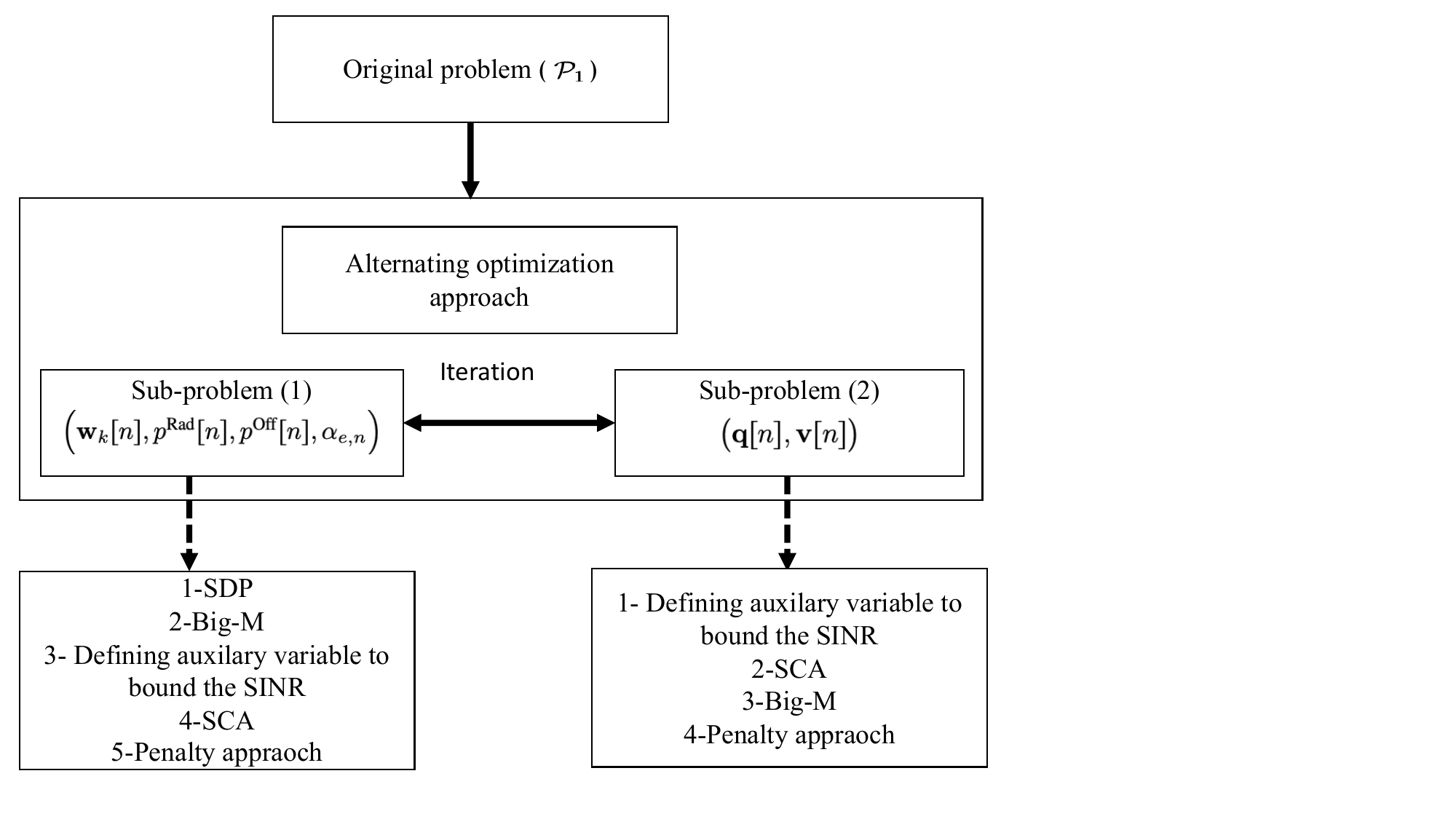}
	\caption{ \small A flow chart of the proposed solution.}\label{flowcahrt}
	\end{figure}
\vspace{-2mm}
\subsection{First Sub-Problem}
  
	First, we assume that the position and velocity of the UAV are fixed, and we aim to optimize the remaining variables. To do so, we employ SDP and define $\mathbf{W}_{k}=\mathbf{w}_{k}\mathbf{w}_{k}^{H} $, $\mathbf{A}_{k}=\mathbf{a}{(\textbf{q}[n],{\textbf{d}}_k)}\mathbf{a}^{H}{(\textbf{q}[n],{\textbf{d}}_k)}$, where $\mathbf{W}_{k}\succeq 0$ and $\text{Rank}(\mathbf{W}_{k})\leq 1$. One obstacle to solving optimization problem $\mathcal{P}_{1}$ is the coupling of $\alpha_{e,n}$ with $\mathbf{W}_{k}[n]$, $\mathbf{R}[n]$, and $p^{\text{Off}}[n]$ in $\text{C1}$-$\text{C4}$.~In order to overcome this difficulty, we adopt the big-M formulation. In particular, we define the new optimization variables,~$\tilde{\mathbf{W}}_{k,e}[n]=\alpha_{e,n}\mathbf{W}_{k}[n]$,  $\tilde{p}^{\text{Rad}}_{e}[n]=\alpha_{e,n}p^{\text{Rad}}[n]$, $\tilde{p}^{\text{Off}}_e[n]=p^{\text{Off}}[n]\alpha_{e,n}$, and add the following additional constraints to the optimization problem:
	\begin{align}
 &\text{C}13:\tilde{p}^{\text{Off}}_{e}[n]\preceq \alpha_{e,n}~P_{\max},~\text{C}14:\tilde{p}^{\text{Off}}_{e}[n]\leq p^{\text{Off}}[n],\\
	&\scalemath{0.9}{\text{C}15:\tilde{p}^{\text{Off}}_{e}[n]\geq 0,~\text{C}16:\tilde{p}^{\text{Off}}_{e}[n]\geq p^{\text{Off}}[n]- (1-\alpha_{e,n})~P_{\max}},
 \end{align}
 \begin{align}
&\text{C}17:\tilde{\mathbf{W}}_{k,e}[n]\preceq \alpha_{e,n}~P_{\max}~\mathbf{I}_{M},\\
	&\text{C}18:\tilde{\mathbf{W}}_{k,e}[n]\preceq \mathbf{W}_{k}[n],~\text{C}19:\tilde{\mathbf{W}}_{k,e}[n]\succeq \mathbf{0},\\
  &\text{C}20:\tilde{\mathbf{W}}_{k,e}[n]\succeq \mathbf{W}_{k}[n]- (1-\alpha_{e,n})~P_{\max}~\mathbf{I}_{M},\\
&\text{C}21:\tilde{p}^{\text{Rad}}_{e}[n]\leq \alpha_{e,n}~P_{\max},\text{C}22:\tilde{p}^{\text{Rad}}_{e}[n]\leq p^{\text{Rad}}[n],\\
	    &\scalemath{0.9}{\text{C}23:\tilde{p}^{\text{Rad}}_{e}[n]\geq 0,~\text{C}24:\tilde{p}^{\text{Rad}}_{e}[n]\leq p^{\text{Rad}}[n]- (1-\alpha_{e,n})~P_{\max}}.
	\end{align}
 Besides, we introduce a set of auxiliary optimization variables $\mu_{k}[n]$ to bound the SINR from below\cite{Rezaei,DISAC} 
\vspace{-2mm}
\begin{equation}\label{mu}
    \scalemath{0.9}{ 0\leq\mu_k[n]\leq \frac{\text{Tr}\big(\mathbf{W}_k[n]\mathbf{A}_k[n]\big)}{\sum_{i\neq k}^{}\text{Tr}\big(\mathbf{W}_i[n]\mathbf{A}_k[n]\big)+\frac{\sigma^2_k}{\beta^{2}_{0}}{\big({{{\left\| {{\mathbf{q}[n]}-{\mathbf{d}_k}} \right\|^2+H^2}}}\big)}}}.
\end{equation}
However, \eqref{mu} is still non-convex. To overcome this issue, by introducing auxiliary variable $\phi_{k}[n]$, we can rewrite $\text{C}2$ as follows:
\vspace{-2mm}
		\begin{align}\label{mui}  &   \text{C2a}:\text{Tr}\big(\mathbf{W}_k[n]\mathbf{A}_k[n]\big)\geq \mu_k[n]\phi_{k}[n],\\
		&\scalemath{0.9}{\text{C2b}: \sum_{i\neq k}^{}\text{Tr}\big(\mathbf{W}_i[n]\mathbf{A}_k[n]\big)+\frac{\sigma^2_k}{\beta^{2}_{0}}{\big({{{\left\| {{\mathbf{q}[n]}-{\mathbf{d}_k}} \right\|^2+H^2}}}\big)}\leq \phi_{k}[n]}.\label{phi}
		\end{align}
		The left-hand side of \eqref{mui} is convex. However, the right-hand side is a product of two terms and is not convex. Nevertheless, we can rewrite the product of the two terms as
		\begin{align}\label{26}
 \hspace*{-2mm}\mu_k[n]\phi_k[n]=\frac{1}{2}\Big[\big(\mu_k[n]+\phi_k[n]\big)^2\hspace*{-0.5mm}-\hspace*{-0.5mm}\big(\mu_k^2[n]+\phi_k^2[n]\big)\Big].
		\end{align}
Note that \eqref{26} is a difference of convex (DC) functions \cite{Ata}.~As a result, the first-order Taylor approximation can be adopted to obtain a convex function and $\mu_k[n]\phi_k[n]$ can be bounded as follows:
	\begin{align}\label{nu}
	  &\mu_k[n]\phi_k[n] \geq 0.5\big(\mu_k[n]+\phi_k[n]\big)^2-\mu_{k}^{(t)}\big(\mu_{k}[n]-\mu_{k}^{(t)}[n]\big)\nonumber\\&\hspace*{18mm}
- \phi_{k}^{(t)}[n]\big(\phi_{k}[n]-\phi_{k}^{(t)}[n]\big)\triangleq \nu_{k}[n],
	\end{align}
where ${t}$ denotes the iteration index for SCA.
Next, we relax the integer constraint and rewrite C11 as follows:
\begin{align}
    &\text{C11a}: 0\leq \alpha_{e,n}\leq 1,~ 
    \text{C11b}: \sum_{e=1}^{E}\sum_{n=1}^{N}\alpha_{e,n}-\alpha^{2}_{e,n}\leq 0. 
\end{align}
Constraint $\text{C11b}$ is a DC function, and we use first-order Taylor approximation to convert the non-convex constraint to the following convex constraint
\begin{align}
&\overline{\text{C11b}}:\sum_{e=1}^{E}\sum_{n=1}^{N}\big(\alpha_{e,n}-\alpha^{(t)}_{e,n}(2\alpha_{e,n}-\alpha^{(t)}_{e,n})\big)\leq 0.
\end{align}
 Now, we introduce a penalty factor $\tau_{1}$ to move constraint $\overline{\text{C11b}}$ to the objective function. $\tau_{1}$ represents the relative importance of recovering binary values for $\alpha_{e,n}$. For a sufficiently large value of $\tau_{1}$, optimization problem $\mathcal{P}_{1}$ is equivalent to the following optimization problem \cite{Penalty}:
\begin{align}
&\mathcal{P}_{2}: \mathop {{\rm{min}}}\limits_{\overline{\boldsymbol{\Xi}}} \frac{1}{N}\sum_{n=1}^{N}\bigg(\eta(\sum_{k=1}^{K} \text{Tr}(\mathbf{W}_k[n])+N_s\frac{t_p}{\delta_{t}}\text{Tr}(\mathbf{R}[n]))\nonumber\\&+\scalemath{0.9}{P_{\text{aero}}(\mathbf{v}[n])+M~P_{\text{static}}+\sum_{e=1}^{E}\alpha_{e,n}\mathcal{P}_{\text{Loc}}+\sum_{e=1}^{E}\underbrace{\alpha_{e,n}p^{\text{Off}}[n]}_{\tilde{p}^{\text{Off}}_{e}[n]}\bigg)}\nonumber\\&+\tau_{1}\bigg(\sum_{e=1}^{E}\sum_{n=1}^{N}\big(\alpha_{e,n}-\alpha^{(t)}_{e,n}(2\alpha_{e,n}-\alpha^{(t)}_{e,n})\big)\bigg)\Bigg)\nonumber\\
	\text{s.t.}~~
& \text{C}1:\sum_{k=1}^{K} \text{Tr}(\mathbf{W}_{k}[n])-\sum_{e=1}^{E}\sum_{k=1}^{K} \text{Tr}(\tilde{\mathbf{W}}_{k,e}[n])+\nonumber\\&\sum_{e=1}^{E}N_s\frac{t_p}{\delta_t}\text{Tr}(\tilde{\mathbf{R}}_{e}[n])+\sum_{e=1}^{E}\tilde{p}^{\text{Off}}_{e}[n] \leq P_{\max},\nonumber\\
	& \text{C2c}:\frac{1}{N}\sum_{n=1}^{N}\log_2(1+\mu_{k}[n])-\nonumber\\&\frac{1}{N}\sum_{n=1}^{N}\sum_{e=1}^{E}\alpha_{e,n}\log_2(1+\mu_{k}[n])\geq R_{\min}^{k},\nonumber\\ 
 & \overline{\text{C2a}}:\text{Tr}\big(\mathbf{W}_k[n]\mathbf{A}_k[n]\big)\geq \nu_{k}[n]\nonumber,\\
& \scalemath{0.9}{\text{C2b}:{\sum_{i\neq k}^{}\text{Tr}\big(\mathbf{W}_i[n]\mathbf{A}_k[n]\big)+\frac{\sigma^2_k}{\beta^{2}_{0}}{\big({{{\left\| {{\mathbf{q}[n]}-{\mathbf{d}_k}} \right\|^2\!\!\!+H^2}}}}}\big)\leq \phi_{k,n}}\nonumber,\\
&\text{C}3:\sum_{n=1}^{N}\frac{\tilde{p}^{\text{Rad}}_{e}[n]\vartheta_{e}\beta_{0}^{2}\mathbf{a}^{H}(\theta_{e}){N_s\frac{t_p}{\delta_t}}{\mathbf{R}}_{d}\mathbf{a}(\theta_{e})}{16\pi \Psi^4_{e}[n]\sigma^{2}_{e}}\geq \text{SNR}_{e}^{\text{th}}\nonumber,\\
 &\text{C}4-\text{C}8,\text{C11a},\text{C}13-\text{C}24,
\end{align}
where $\overline{\boldsymbol{\Xi}}\triangleq\{\mathbf{w}_{k}[n],p^{\text{Rad}}[n],\tilde{p}^{\text{Rad}}_{e}[n],{p}^{\text{Off}}[n], \tilde{p}^{\text{Off}}_{e}[n],\alpha_{e,n},\\ \mu_{k}[n],\phi_{k}[n]\}$.
Constraint $\text{C2c}$ is still non-convex. The non-convexity in constraint $\text{C2c}$ is due to the multiplicative integer variable with continuous value on the left-hand side of the second part of this constraint i.e., $\alpha_{e,n}\log_2(1+\mu_{k}[n])$ and the difference between two logarithm functions. To tackle this constraint, we define \begin{equation}\label{mue}
     \scalemath{0.9}{0\leq\mu_{k,e}[n]\leq \frac{\text{Tr}\big(\tilde{\mathbf{W}}_{k,e}[n]\mathbf{A}_k[n]\big)}{\sum_{i\neq k}^{}\text{Tr}\big(\mathbf{W}_{i}[n]\mathbf{A}_k[n]\big)+\frac{\sigma^2_k[n]}{\beta^{2}_{0}}{\big({{{\left\| {{\mathbf{q}[n]}-{\mathbf{d}_k}} \right\|^2+H^2}}}\big)}}}.
\end{equation}
Now, by exploiting auxiliary variable $\phi_{k}[n]$, we can rewrite \eqref{mue} as follows:
\vspace{-2mm}
		\begin{align}\label{muie}  &   \text{C2d}:\text{Tr}\big(\tilde{\mathbf{W}}_{k,e}[n]\mathbf{A}_k[n]\big)\geq \mu_{k,e}[n]\phi_{k}[n].
		\end{align}
The left-hand side of \eqref{muie} is convex. However, the right-hand side is a product of two terms and is not convex. We treat this constraint in a similar way as \eqref{26} and \eqref{nu}. As a result, constraint C2c can be rewritten as 
\begin{align}\label{C2C}
\scalemath{0.9}{\frac{1}{N}\sum_{n=1}^{N}\log_2(1+\mu_{k}[n])-\frac{1}{N}\sum_{n=1}^{N}\sum_{e=1}^{E}\log_2(1+\mu_{k,e}[n])\geq R_{\min}^{k}}.
\end{align}
The left-hand side of \eqref{C2C} is a difference of two concave functions which is not generally concave. Hence, we employ a first order Taylor approximation to obtain a concave function, i.e., $\log_2(1+\mu_{k,e}[n])$ is bounded as follows:
\begin{align}\label{muke}
&f(\mu_{k,e}[n])\triangleq\log_2(1+\mu_{k,e}[n])\leq \log_2(1+\mu^{(t)}_{k,e}[n])\nonumber\\&+\frac{\partial f(\mu_{k,e}[n])}{\partial \mu_{k,e}[n]}\big(\mu_{k,e}[n]-\mu_{k,e}^{(t)}[n]\big)\triangleq \tilde{f}(\mu_{k,e}[n]).
\end{align}
The next challenge is addressing the requirements of equality constraint C12. C12 enforces the exact alignment between the UAV and the specified target position, which may slow down the speed of convergence of the proposed iterative algorithm. To overcome this limitation, we introduce a penalty function to relax the strictness of the constraint during the iterative process, making it more flexible and facilitating convergence. Following the principles of the penalty method, we modify C12 to $\overline{\text{C}12}$, i.e., $\overline{\text{C}12}:\alpha_{e,n}\big\|\mathbf{q}[n]-\mathbf{d}_{e}\big\|^{2}\leq 0$ and introduce a penalty function into the objective function, penalizing constraint violations with a coefficient $\tau_{2} > 0$ \cite{Penalty}. As a result, the optimization problem at hand can be rewritten as 
\begin{align}
&\mathcal{P}_{3}: \mathop {{\rm{min}}}\limits_{\widetilde{\mathbf{\Xi}}} \frac{1}{N}\sum_{n=1}^{N}\bigg(\eta(\sum_{k=1}^{K} \text{Tr}(\mathbf{W}_k[n])+N_s\frac{t_p}{\delta_{t}}p^{\text{Rad}}[n]+\nonumber\\&\scalemath{0.9}{P_{\text{aero}}(\mathbf{v}[n])+M~P_{\text{static}}+\sum_{e=1}^{E}\alpha_{e,n}\mathcal{P}_{\text{Loc}}[n]+\sum_{e=1}^{E}\underbrace{\alpha_{e,n}p_{\text{Off}}[n]}_{\tilde{p}^{e}_{\text{Off}}[n]}\bigg)}+\nonumber\\&\tau_{1}\bigg(\sum_{e=1}^{E}\sum_{n=1}^{N}\big(\alpha_{e,n}-\alpha^{(t)}_{e,n}(2\alpha_{e,n}-\alpha^{(t)}_{e,n})\big)\bigg)\Bigg)\nonumber+\\&\tau_{2}\bigg(\sum_{e=1}^{E}\sum_{n=1}^{N}\alpha_{e,n}\big\|\mathbf{q}[n]-\mathbf{d}_{e}\big\|^{2}\bigg)\nonumber\\
	\text{s.t.}~~
& \text{C}1:\sum_{k=1}^{K} \text{Tr}(\mathbf{W}_{k}[n])-\sum_{e=1}^{E}\sum_{k=1}^{K} \text{Tr}(\tilde{\mathbf{W}}_{k,e}[n])+\nonumber\\&\sum_{e=1}^{E}N_s\frac{t_p}{\delta_t}\tilde{p}_{e}^{\text{Rad}}[n]+\sum_{e=1}^{E}\tilde{p}_{e}^{\text{Off}}[n] \leq P_{\max},\label{limit_power}\nonumber\\
	&\scalemath{0.95}{\overline{\text{C2c}}:\frac{1}{N}\!\!\sum_{n=1}^{N}\log_2(1+\mu_{k}[n])-\!\!\frac{1}{N}\!\!\sum_{n=1}^{N}\sum_{e=1}^{E}\tilde{f}(\mu_{k,e}[n])\geq R_{\min}^{k}},\nonumber\end{align}
\begin{align}
&\overline{\text{C2a}}:\text{Tr}\big(\mathbf{W}_k[n]\mathbf{A}_k[n]\big)\geq \nu_{k}[n],~\text{C2b},\nonumber\\&\overline{\text{C2d}}:\text{Tr}\big(\tilde{\mathbf{W}}_{k,e}[n]\mathbf{A}_k[n]\big)\geq \nu_{k,e}[n],\nonumber\\
&\text{C}3:\sum_{n=1}^{N}\frac{\tilde{p}^{\text{Rad}}_{e}[n]\vartheta_{e}\beta_{0}^{2}\mathbf{a}^{H}(\theta_{e}){N_s\frac{t_p}{\delta_t}}{\mathbf{R}}_{d}\mathbf{a}(\theta_{e})}{16\pi \Psi^4_{e}[n]\sigma^{2}_{e}}\geq \text{SNR}_{e}^{\text{th}},\nonumber\\
 &\text{C}4-\text{C}8,\text{C11a},\text{C13}-\text{C}24,
\end{align}
where $\widetilde{\mathbf{\Xi}}=\{\mathbf{W}_{k}[n],\tilde{\mathbf{W}}_{k,e}[n],p^{\text{Rad}}[n],\tilde{p}_{e}^{\text{Rad}}[n],\tilde{p}_{e}^{\text{Off}}[n],p^{\text{Off}}[n],\\\alpha_{e,n},\mu_{k}[n],\mu_{k,e}[n],\phi_{k}[n]\}$ is the new set of optimization variables.
Now, by dropping the rank-one constraint on $\mathbf{W}_{k}[n]$ and adopting SDP relaxation, problem $ \mathcal{P}_3 $ becomes a convex optimization problem and can be efficiently solved by CVX. The tightness of the SDP relaxation can be proved following similar steps as in \cite[Appendix A]{Rank}. We omit the proof here due to lack of space.
\subsection{Second Sub-Problem }   
	In the subsequent step of our proposed solution, the trajectory and velocity of the UAV are designed. Optimal trajectory design poses a challenge as the UAV's position affects the steering verctor, rendering the problem intractable. Additionally, the complexity is increased by the non-convex nature of the data rate constraint in C2, a pivotal component of our optimization problem. Despite these challenges, we derive a high-quality suboptimal solution. To this end, we introduce new auxiliary optimization variables $\beta_{k}[n]$ and $\mu'_{k}[n]$ to effectively bound the SINR. This transformation allows us to reframe $\text{C2}$ into a set of equivalent constraints. Consequently, $\text{C2}$ is equivalently replaced by the following constraints
		\begin{align}
&\widehat{\text{C2a}}: \text{Tr}\big(\mathbf{W}_k[n]\widetilde{\mathbf{H}}_k[n]\big)\geq \mu'_{k}[n]\beta_{k}[n],\label{mu'1}\\
&\widehat{\text{C2b}}:\!\!\!\sum_{i\neq k}^{}\!\text{Tr}\big(\mathbf{W}_i[n]\widetilde{\mathbf{H}}_k[n]\big)+\sigma^2_k(\|\mathbf{q}[n]-{\mathbf{d}}_k\|^2+H^2)\leq \beta_k[n]\label{mu'2},
		\end{align}
  where $ \widetilde{\mathbf{H}}_k[n]= \beta_0^2\mathbf{A}_{k}[n]$.
The right-hand side of \eqref{mu'1} is not a convex function. Similarly as in \eqref{nu}, by adopting the first-order Taylor approximation we obtain a convex function as $\chi_k[n] \triangleq 0.5\big(\mu^{\prime}_k[n]+\beta_k[n]\big)^2-
\mu_{k}^{\prime(t)}\big(\mu^{\prime}_{k}[n]-\mu_{k}^{\prime(t)}[n]\big)- \beta_{k}^{(t)}[n]\big(\beta_{k}[n]-\beta_{k}^{(t)}[n]\big)$, where $t$ denotes again the SCA iteration index. The left-hand side of \eqref{mu'1} is also a non-convex function of the UAV's position $\mathbf{q}[n]$. 
Nevertheless, we can rewrite the left-hand side of \eqref{mu'1} as follows
\begin{align}
&\text{Tr}\big(\mathbf{W}_k[n]\widetilde{\textbf{H}}_k[n]\big)=\beta_0^2\sum_{m=1}^{M}\!\sum_{m'=1}^{M}\!\!\mathbf{W}^k_{m,m'}[n]e^{\frac{j2\pi\frac{\hat{d}}{\lambda}H (m'-m)}{\sqrt{\|\textbf{q}[n]-{\textbf{d}}_k\|^2+H^2}}}\nonumber\\=&\underbrace{\beta_0^2 \sum_{m=1}^{M}\mathbf{W}^k_{m,m}[n]}_{\triangleq U_k[n](\mathbf{W}_{k})}+\beta_0^2 \sum_{m=1}^{M}\sum_{m'=m+1}^{M}|\mathbf{W}^k_{m,m'}[n]|\nonumber\\&\cos \bigg(2\pi\frac{\hat{d}}{\lambda}(m'-m)\frac{H}{\sqrt{\|\mathbf{q}[n]-{\textbf{d}}_k\|^2+H^2}}+\phi^{{W}_{k}}_{m,m'}[n]\bigg)\!\!\triangleq\nonumber\\&\! U_k[n](\mathbf{W}_{k})+ J_k[n](\mathbf{W}_{k},\mathbf{q})\label{J},
 \end{align}
	where $ \mathbf{W}_{m,m'}^k[n]$ is the element in the $ m^{\text{th}} $ row and  $ m'^{{\text{th}}}$ column of $ \textbf{W}_k[n]$. Besides, $|\mathbf{W}^k_{m,m'}[n]|$ and $ \phi^{{W}_{k}}_{m,m'}[n]$ denote the magnitude and phase of $ \mathbf{W}^k_{m,m'}[n]$, respectively. 
 Note that since the right-hand side of \eqref{mu'1} is convex, we need 
	 to find an affine approximation of $J_k[n]$ to convexify the underlying optimization problem, which is done via a first-order Taylor approximation as follows
		\begin{equation}
		 \tilde{J}_k[n](\mathbf{W}_{k},\mathbf{q})\triangleq J^{(t)}_k[n](\mathbf{W}_{k},\mathbf{q})+\boldsymbol{\nabla}^{H}_{{J}_k[n]}\big(\mathbf{q}[n]-\mathbf{q}^{(t)}[n]\big),\label{J_tilde}
		\end{equation}
where gradient $\boldsymbol{\nabla}_{{J}_k[n]}$ is given by 
\begin{align}
& \scalemath{0.9}{\boldsymbol{\nabla}_{{J}_k[n]}=\dfrac{-4\beta_0^2\pi\hat{d}H(m'-m)}{\lambda\big(\sqrt{\|\mathbf{q}^{(t)}[n]-{\mathbf{d}}_k\|^2+H^2\big)}^{3}} \sum_{m=1}^{M}\sum_{m'=m+1}^{M}|\mathbf{W}^k_{m,m'}[n]|}\nonumber\\&\sin \bigg(2\pi\frac{\hat{d}}{\lambda}(m'-m)\frac{H}{\sqrt{\|\mathbf{q}^{(t)}[n]-{\mathbf{d}}_k\|^2+H^2}}+\phi^{{W}_{k}}_{m,m'}[n]\bigg)\nonumber\\&(\mathbf{q}^{(t)}[n]-{\mathbf{d}}_k).
\end{align}	
By substituting \eqref{J_tilde}, \eqref{mu'1} can be restated as follows
\begin{align}
\widehat{\overline{\text{C2a}}}:~{U}_k[n](\mathbf{W}_{k})+ \tilde{J}_k[n](\mathbf{W}_{k},\mathbf{q})\geq \chi_{k}[n]. \label{40}
\end{align}
Similarly, the left-hand side of \eqref{mu'2} can be approximated by its first-order Taylor series. As a result, the inequality in \eqref{mu'2} can be restated as
	\begin{align}
	&\widehat{\overline{\text{C2b}}}:\sum_{i\neq k}^{}\big(U_i[n](\mathbf{W}_{i}) + \tilde{J}_{i}[n](\mathbf{W}_{i},\mathbf{q})\big)+\nonumber\\& \sigma^{2}_{k}({\|\mathbf{q}[n]-{\mathbf{d}}_k\|^2+H^2})\leq \beta_{k}[n]\label{R}.
	\end{align}

	Finally, we deal with the non-convexity of the power consumption of a moving UAV. To do so, we introduce the auxiliary variable $y[n]\geq 0$, such that 
	\begin{align}
	    y^{2}[n]=\sqrt{1+\frac{\|\mathbf{v}[n]\|^4}{4 v_0^4}}-\frac{\|\mathbf{v}[n]\|^2}{2v_0^2},
	\end{align}
	which can be rewritten as 
  \vspace{-2mm}
	\begin{align}\label{EnU}
	    \frac{1}{y^{2}[n]}=y^{2}[n]+\frac{\|\mathbf{v}[n]\|^2}{v_0^2}.
	\end{align}
	Consequently, the second term in the aerodynamic power consumption during UAV flight can be restated as $P_i\big(y(n)-1\big)$.~Hence, the total aerodynamic power consumption during UAV flight can be restated as  $\tilde{P}_{\text{fly}}$=$
P_o \bigg(\frac{3 \|\mathbf{v}[n]\|^2 }{\Omega^2 r^2} \bigg) + P_i\big(y(n)-1\big)+ \frac{1}{2} r_0 \rho s A_{\mathrm r} \|\mathbf{v}[n]\|^3$.
With the above manipulations, the optimization problem is recast as follows
\vspace{-2mm}
\begin{align}
	 &\mathcal{P}_{5}: \mathop {{\rm{min}}} \limits_{\scriptstyle{ \mathbf{q}},\mathbf{v},y,\mu'_{k},\beta_{k}}\mathcal{F}\triangleq\frac{1}{N}\sum_{n=1}^{N}\bigg(\sum_{e=1}^{E}\alpha_{e,n}P_{\text{hover}}[n]\nonumber+\\&\scalemath{0.9}{(1-\sum_{e=1}^{E}\alpha_{e,n})\tilde{P}_{\text{fly}}(\mathbf{v}[n])\bigg)+\tau_{2}\bigg(\sum_{e=1}^{E}\sum_{n=1}^{N}\alpha_{e,n}\big\|\mathbf{q}[n]-\mathbf{d}_{e}\big\|^{2}\bigg)}\nonumber\\
	\text{s.t.}~~
	&\scalemath{0.85}{\text{C2c}:\frac{1}{N}\sum_{n=1}^{N}\log_2(1+\mu'_{k}[n])-\frac{1}{N}\sum_{n=1}^{N}\sum_{e=1}^{E}\log_2(1+\mu'_{k,e}[n])\geq R_{\min}^{k}},\nonumber\\
 &\text{C}4: \log_2\Big(1+\frac{\alpha_{e,n}p^{\text{Off}}[n]\beta _0^{2} G_{T}}{(\|\mathbf{q}[n]-{\mathbf{q}}_b\|^2+H_{b}^2)\sigma^{2}_{B}}\Big)\geq \alpha_{e,n}\iota R_{\text{Pr}},\nonumber\\
 &\text{C5}:\log_2\Big(1+\frac{p_{\text{BS}}[n]\beta _0^{2} G_{T}}{(\|\mathbf{q}[n]-{\mathbf{q}}_b\|^2+H_{b}^2)\sigma^{2}}\Big)\geq \nonumber\\&\sum_{k=1}^{  K}R^{k}_{\min}(1-\sum_{e=1}^{E}\alpha_{e,n}),~\nonumber\\
 &\text{C26}:\frac{1}{y^{2}[n]}\leq y^{2}[n]+\frac{\|\mathbf{v}[n]\|^2}{v_0^2}\nonumber,\\~
    &\widehat{\overline{\text{C2a}}}, \widehat{\overline{\text{C2b}}},\text{C9}-\text{C11}.
	\end{align}
 Problem $\mathcal{P}_{5}$ is still non-convex due to non-convex constraints C2c, C4, C5, and C26. We address C2c by applying a similar approach as in \eqref{muke}. Moving on to C4, this constraint can be equivalently restated as follows:
\begin{align}
	&\overline{\text{C}4}: \|\mathbf{q}[n]-{\mathbf{q}}_b\|^2+H_{b}^2\leq \nonumber\\ &\frac{p^{\text{Off}}[n]\beta_{0}^{2} G_{T}}{\sigma^{2}_{B}}\bigg[\frac{\alpha_{e,n}}{2^{(\iota R_{\text{Pr}})}-1}+(1-\alpha_{e,n})\mathcal{M}\bigg],
\end{align}
where $\mathcal{M}$ represents a sufficiently large value. The value of $\mathcal{M}$ should be chosen large enough to ensure that the constraint is always fulfilled for $\alpha_{e,n}=0$. Finally, SCA can be used to effectively handle constraint C26 by deriving a corresponding global
lower bound at a given local point. As a result, based on the first-order Taylor approximation of the right-hand side of \text{C26}, the following global lower bound can be obtained
\begin{align}
&y^{2}[n]+\frac{\|\mathbf{v}[n]\|^2}{v_0^2}\geq y^{(t)2}{[n]}+\frac{\|\mathbf{v}^{(t)}[n]\|^2}{v_0^2}+\nonumber\\&2y^{(t)}{[n]}(y{[n]}-y^{(t)}{[n]})+\frac{2\mathbf{v}^{(t)}[n]}{v_{0}^{2}}(\mathbf{v}[n]-\mathbf{v}^{(t)}[n])\nonumber\\&\triangleq g(y[n],\mathbf{v}[n]),
\end{align}
where $y^{(t)}[n]$ and $\mathbf{v}^{(t)}[n]$ are the values obtained in the $t$-th iteration of SCA. 
This leads to the following convex optimization problem 
\begin{align}
  \mathcal{P}_{6}: &\mathop {{\rm{min}}} \limits_{\scriptstyle{ \mathbf{q}},\mathbf{v},y,\mu'_{k},\beta_{k}}\mathcal{F}\nonumber\\
	\text{s.t.}~&\widetilde{\text{C}26}:\frac{1}{y^{2}[n]}\leq g(y[n],\mathbf{v}[n])\nonumber,\\
	&\overline{\text{C}4}: \|\mathbf{q}[n]-{\mathbf{q}}_b\|^2+H_{b}^2\leq \nonumber\\ &\frac{p^{\text{Off}}[n]\beta _0^{2} G_{T}}{\sigma^{2}_{B}}\bigg[\frac{\alpha_{e,n}}{2^{(\iota R_{\text{Pr}})}-1}+(1-\alpha_{e,n})\mathcal{M}\bigg],\nonumber\\ &	\overline{\text{C5}}:\|\mathbf{q}[n]-{\mathbf{q}}_b\|^2+H_{b}^2\leq\nonumber\\ & \frac{p_{\text{BS}}[n]\beta _0^{2} G_{T}}{\bigg(2^{\big(\sum_{k=1}^{  K}R^{k}_{\min}(1-\sum_{e=1}^{E}\alpha_{e,n})\big)}-1\bigg)\sigma^{2}},~\nonumber\\&\overline{\text{C2c}},\widehat{\overline{\text{C2a}}},\widehat{\overline{\text{C2b}}},\text{C9}-\text{C11}.
\end{align}
In each iteration $t$, we update the solution set and efficiently solve $\mathcal{P}_{6}$ via CVX.
\subsection{Overall Algorithm}
The proposed solution based on AO is summarized in \textbf{Algorithm 1}.  In the following, we analyze the convergence and complexity of the proposed algorithm. 

\subsubsection{Convergence}The proposed AO algorithm optimizes two sets of variables iteratively: (i) $\boldsymbol{\Pi} \triangleq \{\mathbf{w}_{k}[n], p^{\text{Rad}}[n], p^{\text{Off}}[n], \alpha_{e,n}\}$ encompassing the beamforming vectors, radar power, offloading power, and sensing indicator variables, and (ii) $\{\mathbf{v}, \mathbf{q}\}$ representing the UAV's velocity and trajectory. The algorithm employs SCA and SDR to overcome the non-convexity of the formulated optimization problem. Specifically, when set $\boldsymbol{\Pi}$ is updated while $\mathbf{v}$ and $\mathbf{q}$ are fixed, we solve $\mathcal{P}_{3}$ that ensures $\mathcal{O}bj(\boldsymbol{\Pi}^{(t+1)},{v}^{(t)}, \mathbf{q}^{(t)}) \leq \mathcal{O}bj(\boldsymbol{\Pi}^{(t)}, \mathbf{v}^{(t)}, \mathbf{q}^{(t)})$. Similarly, updating $\mathbf{v}$ and $\mathbf{q}$ while fixing $\boldsymbol{\Pi}^{(t+1)}$ results in
$\mathcal{O}bj(\boldsymbol{\Pi}^{(t+1)}, \mathbf{v}^{(t+1)}, \mathbf{q}^{(t+1)}) \leq \mathcal{O}bj(\boldsymbol{\Pi}^{(t+1)}, \mathbf{v}^{(t)}, \mathbf{q}^{(t)})$. Moreover, for sufficiently large penalty factors $\tau_{i}$, $i \in \{1,2\}$, in problems $\mathcal{P}_{3}$ and $\mathcal{P}_{6}$, the objective function of $\mathcal{P}_{1}$ is non-increasing in each iteration of \textbf{Algorithm 1}  \cite{Xu_iter, tseng2001convergence} and converges to a stationary value of the objective function in $\mathcal{P}_{1}$, producing a high-quality suboptimal solution\cite{Ata_ICC}. 

\subsubsection{Computational Complexity}The computational complexity of \textbf{Algorithm 1} is given by $\mathcal{O}\Big(\mathrm{log}(1/\varepsilon_{\text{AO}})\big((3N+K+3NK+4EKN+10EN+2E)M^{3}+(3N+K+3NK+4EKN+10EN+2E)^{2}M^2+(5N+3NK+K+2NE)(M)^3+(5N+3NK+K+2NE)^{2}M^2\Big)$, where $\mathcal{O}\left ( \cdot  \right )$ is the big-O notation and $\varepsilon_{\text{AO}}$ is the convergence tolerance of \textbf{Algorithm 1}.

\begin{algorithm}[t]
    \footnotesize
    \captionof{algorithm}{Proposed resource allocation framework.}
     \label{algorithm}
     1.\quad Initialize $\alpha_{e,n}^{(t)}$,~$\mathbf{v}^{(t)}[n]$,~$\mathbf{q}^{(t)}[n]$ ,~$\mu_k^{(t)}$,~$\phi_k^{(t)}$, $\beta_k^{(t)}$, $\mu_k^{\prime(t)}$, $\tau_{1,2} \gg 1$, $t$~ \text{(iteration index)},~$\varepsilon_{\text{AO}}$.\\
     \textbf{Repeat} \\
  2.\quad Solve $\mathcal{P}_3$ for given $\mathbf{v}[n]=\mathbf{v}^{(t)}[n]$, $\mathbf{q}[n]=\mathbf{q}^{(t)}[n]$ and obtain $\mathbf{W}_{k}^{(t+1)}[n]$, $\mathbf{R}^{(t+1)}[n]$, $p^{\text{Off}~(t+1)}[n]$ and $\alpha_{e,n}^{(t+1)}$.\\
3. \quad Solve $\mathcal{P}_6$ for given $\mathbf{W}_{k}[n]= \mathbf{W}_{k}^{(t+1)}[n]$, $\mathbf{R}[n]= \mathbf{R}^{(t+1)}[n]$, $\alpha^{(t+1)}_{e,n}$, and obtain $\mathbf{v}^{(t+1)}[n]$,~$\mathbf{q}^{(t+1)}[n]$.\\
5. \quad Set~$t=t+1$\\
       6. \quad \textbf{until}~$\frac{\mathcal{O}bj^{(t)}-\mathcal{O}bj^{(t-1)}}{\mathcal{O}bj^{(t-1)}}\leq \varepsilon_{\text{AO}}$.\\ 
            \vspace{-1mm}
     \end{algorithm}
\begin{table}[t]\vspace*{-2mm} \caption{System simulation parameters.}\label{ISAC_parameters}\footnotesize
\newcommand{\tabincell}[2]{\begin{tabular}{@{}#1@{}}#2\end{tabular}}
\centering
\begin{tabular}{|l|l|l|}
\hline
    \hspace*{-1mm}$\sigma_{e}^{2}=\sigma_{k}^2$& Noise power & $-110$ dBm \\
\hline
    \hspace*{-1mm}$T$& Time horizon & $70$ s \\
\hline
\hspace*{-1mm}$\delta_{t}$& Duration of one time slot & $1$ s \\
\hline
    \hspace*{-1mm}$P_{\mathrm{max}}$& Maximum transmit power at the UAV & $40$ dBm\\
\hline
    \hspace*{-1mm}${R}_{\min}$&  Required achievable rate of users & $1$ bits/s/Hz \\
\hline
\hspace*{-1mm}${\vartheta}_{e}$&  RCS & $0.1$~m$^{2}$ \\
\hline 
    \hspace*{-1mm}$\text{SNR}_{e}^{\text{th}}$& Minimum long-term sensing SNR at the UAV & $5$ dB \\
\hline
    \hspace*{-1mm}$M$& Number of antennas at the UAV & $6$ \\
\hline
\hspace*{-1mm}$G_{T}$& Antenna gain & $10$~dBi \\
\hline
    \hspace*{-1mm}$R_{\text{Pr}}$& Fronthaul link capacity (rate of production) & $4$ bits/s/Hz \\
\hline
    \hspace*{-1mm}$N_b$&  Number of bits for quantizing echoes & $4$ \\
\hline
    \hspace*{-1mm}$N_{s}$&  Number of rounds in sensing phase & $4400$ \\
\hline
    \hspace*{-1mm}$\Delta R$ &  Sensing resolution  & $15$ m \cite{radarUAV}\\
\hline
    \hspace*{-1mm}$W_{\mathrm{F}}$ & Fronthaul link bandwidth & $10$ MHz  \cite{radarUAV}\\
\hline
    \hspace*{-1mm}$t_{p}$ &  Pulse width & $0.6~\mu$s  \\
\hline
\hspace*{-1mm}$t_{o}$ &  Listening time & $2.26*10^{-4}$s \\
\hline 
\hspace*{-1mm}$2\Delta$ & Beamwidth of the ideal beam pattern & $\frac{\pi}{6}$ \\
\hline
    \hspace*{-1mm}$\epsilon_{\text{AO}}$ & Convergence tolerance & $10^{-3}$ \\
\hline
    \hspace*{-1mm}$\tau_{\{1,2\}}$ & Penalty factors & $10^{5}$ \\
\hline
\hspace*{-1mm} $a_{\max}$ & UAV maximum acceleration & $5$ $\text{m/s}^{2}$ \\
\hline
\hspace*{-1mm} PRF & Pulse repetition frequency & $4.4$ kHz \cite{radarUAV}\\
\hline
\hspace*{-1mm} $a$ & Hardware architecture & $10^{-28}$ \cite{MEC_UAV}\\
\hline
$f_{\text{Loc}}$ & Local computation resource at the UAV & $3$ GHz \cite{MEC_UAV}\\
\hline
\end{tabular}
\end{table}
\vspace{-7mm}
\section{Simulation Results}
\vspace{-2mm}
In this section, we evaluate the performance of the proposed algorithm via computer simulations.~We consider an area of $0.3$ km $\times$ $0.3$ km with $K=3$ communication users and $E=3$ sensing targets. The UAV is equipped with $M=6$ antennas and we set the minimum long-term sensing SNR at the UAV to $\text{SNR}_{e}^{\text{th}}=5$~dB \cite{ThUAVISAC}.~ Furthermore, the UAV operates at an altitude of $H=100$ meters\footnote{Operating at an altitude of 100 meters enhances the probability that the UAV maintains an LoS connection to the communication user, as illustrated by the relationship between the altitude and the LoS probability in \cite{LoS}. We note that, in practice, if a reliable LoS link cannot be established due to blockages or unfavorable user locations the affected users are not admitted into the system. Admission control is handled by a corresponding protocol\cite{3gpp5g}. This protocol is crucial for maintaining system integrity and operational efficiency but falls outside the scope of this paper. Thus, in this paper, we consider only users, which have been admitted into the system to participate in communication.} with a maximum flight speed of $v_{\max}=15$ m/s. Additionally, the channel power gain at the reference distance of $d_{0}=1$ meter is $\beta_0=-30$ dB. The adopted simulation parameters are given in Table II. To investigate the power saving achieved by the proposed scheme, we compare it with two baseline schemes. For baseline scheme 1, we adopt a heuristic trajectory design for the UAV. In particular, the UAV visits each communication user and sensing target along the shortest path while the beamformers, sensing power, sensing indicator, and velocity are optimized. For baseline scheme 2, we adopt zero-forcing beamforming for information transmission and a fixed velocity i.e., $v_{\text{fixed}}=13$ m/s, omitting constraint C9. Then, we jointly optimize the communication and sensing power, sensing indicator, and trajectory using a modified version of $\mathcal{P}_{1}$.
\begin{figure*}[t]
	\centering
	\vspace{-3mm}
	\subfigure[UAV trajectory for BS position (a).]{
		\label{figure3a}
		\vspace{-3mm}
		\includegraphics[width=7.6cm]{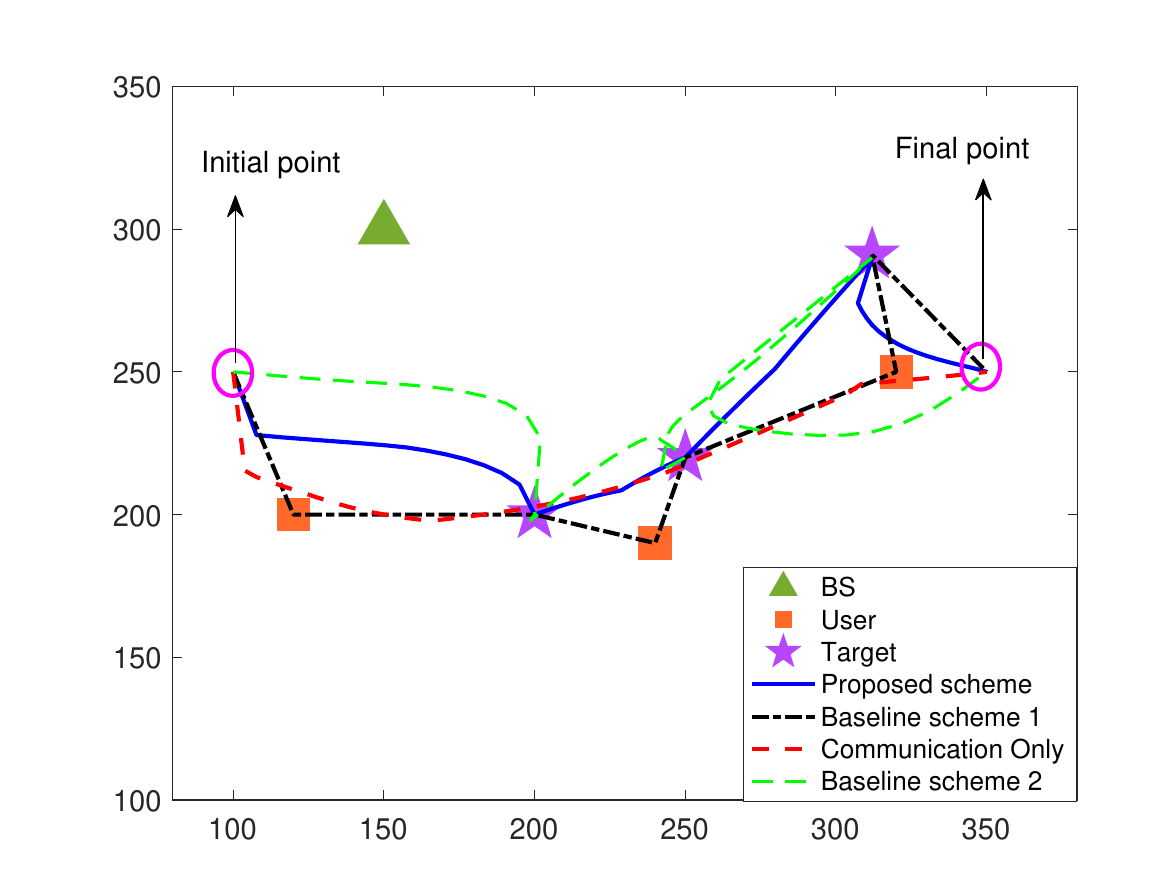}
	}
	\subfigure[UAV trajectory for BS position (b).]{
		\label{figure3b}
		\includegraphics[width=7.6cm]{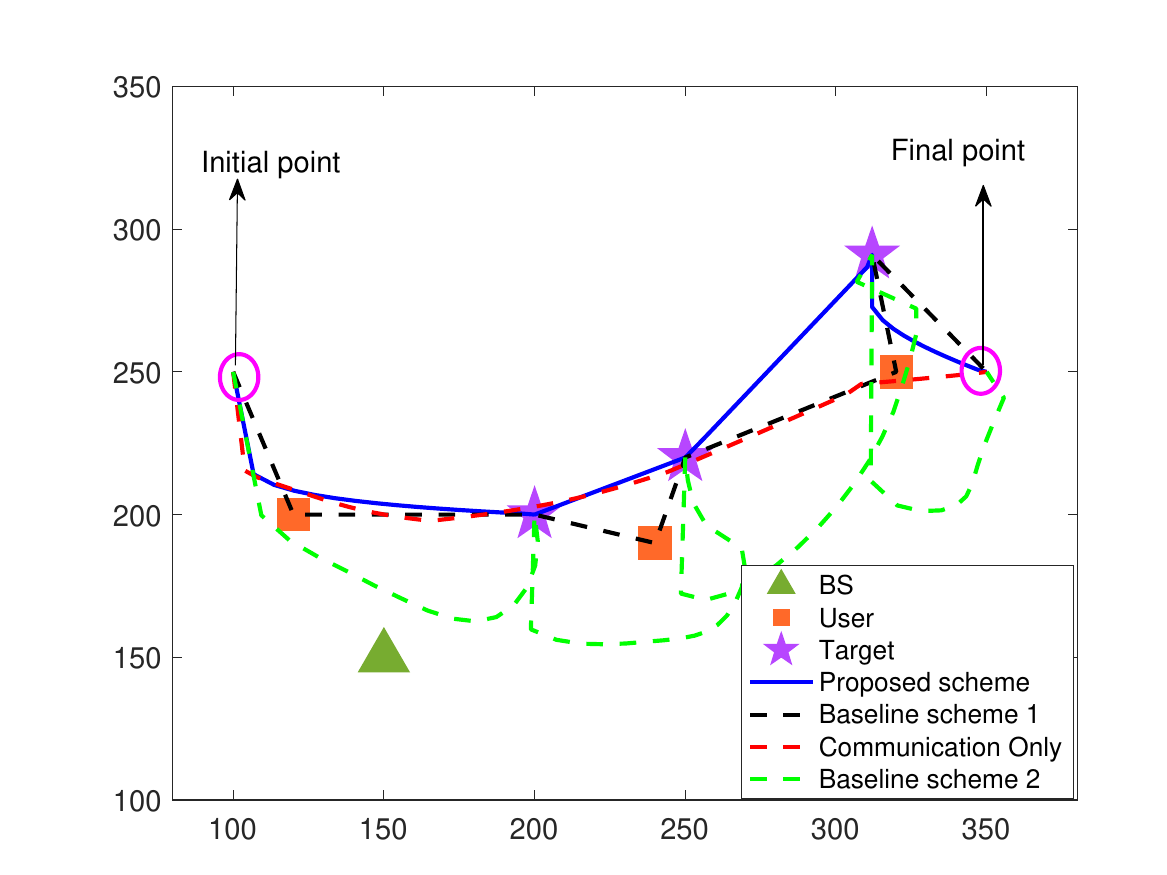}
	}
 \subfigure[Velocity of the UAV versus time for BS location (a).]{
		\label{figure3c}
		\includegraphics[ height= 5.500cm,width=6.8cm]{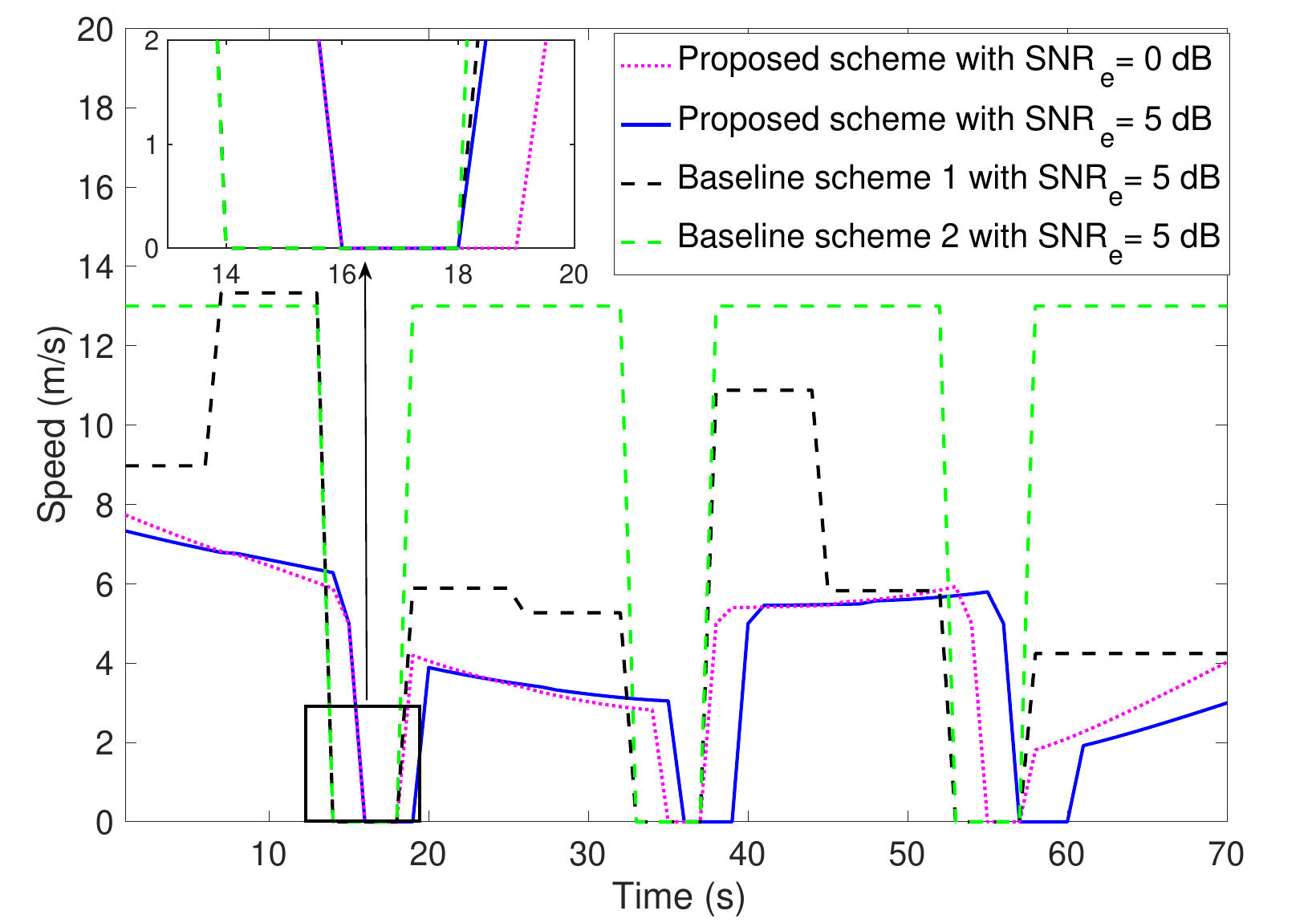}
	}
 \subfigure[Aerodynamic power consumption of the UAV versus time for BS location (a).]{
		\label{figure3d}
		\includegraphics[height= 5.700cm,width=7.6cm]{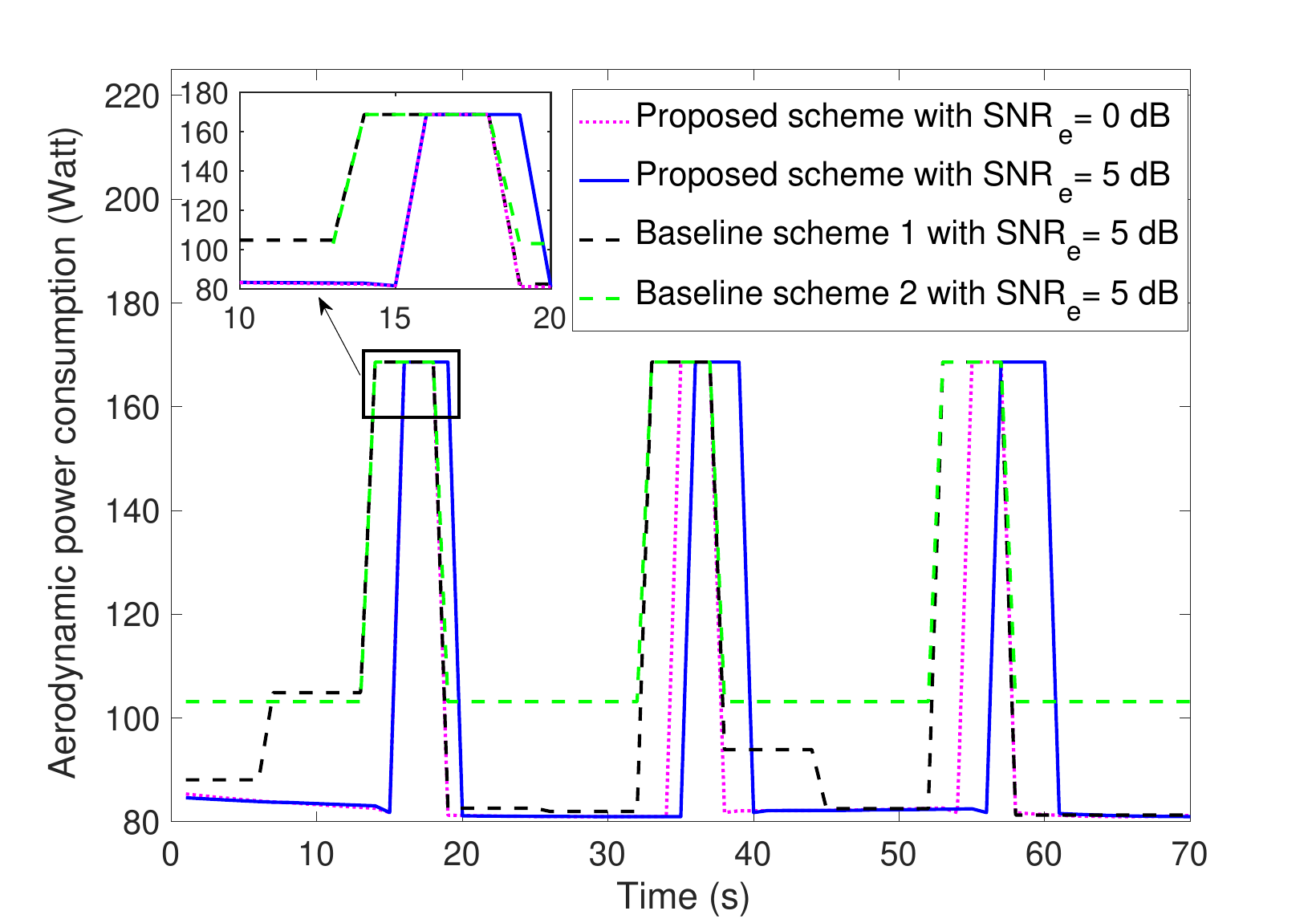}
	}
	\caption {Trajectory, velocity, and aerodynamic power consumption of the UAV.}
	\label{figure6}
	\vspace{-3mm}
\end{figure*} 

Figs. 4(a) and 4(b) depict the trajectory of the UAV during its mission for different positions of the ground BS. In particular, for the proposed scheme, the UAV starts flying from the initial point towards the location of the first sensing target while transmitting data to the communication users. 
Fig.~4(c) shows that, during this time, the UAV also adjusts its velocity to minimize power consumption. When approaching the first sensing target, the UAV gradually reduces its velocity to zero before hovering above the target for sensing. Subsequently, the UAV proceeds to the second target, executing another hover-and-sense operation. This pattern continues as the UAV navigates to the third target. In the intervals between the target locations, the UAV dynamically adjusts its trajectory and beamforming vectors to ensure continuous communication, while meeting the data rate requirements of the communication users.
 As the mission nears completion, the UAV heads towards its final destination while continuing to support the communication users. It is worth noting that the UAV's trajectory exhibits curvature. This is because in order to save power, the UAV aims to fly at the optimal velocity, while properly adjusting its distance to both the users and the BS for efficient information transmission. Figs. 4(a) and 4(b) also include the trajectories of the UAV when there is no sensing requirement. In this case, the UAV conserves power by navigating between the users, while supporting multiple users simultaneously.~From Fig. 4(c), we can observe that for baseline scheme 1, as the trajectory is not optimized, the UAV needs to fly with a higher velocity to complete its mission. This results in increased aerodynamic power consumption, as illustrated in Fig. 4(d).~Another interesting observation is that the proposed algorithm leads to shorter hovering times compared to baseline schemes 1 and 2. Because of the joint optimization of the sensing indicator, beamformers for information transmission, sensing power, trajectory, and  UAV velocity, less time is needed to complete the sensing tasks. Moreover, as the sensing requirements become more stringent, the required UAV hovering time increases. This increase in hovering time, in turn, leads to a corresponding rise in the aerodynamic power consumption of the UAV. A comparison of Figs. 4(a) and 4(b) reveals the influence of the location of the ground BS on the UAV's trajectory. In fact, the UAV strives to maintain close proximity to the BS throughout its mission. This behavior is caused by constraints C4 and C5, which ensure real-time information exchange between the UAV and BS. Consequently, during each time slot, the UAV flies as close as possible to the ground BS, enhancing the reliability of the connection. 



\begin{figure}[t]
    \centering
\includegraphics[width=8.6cm]{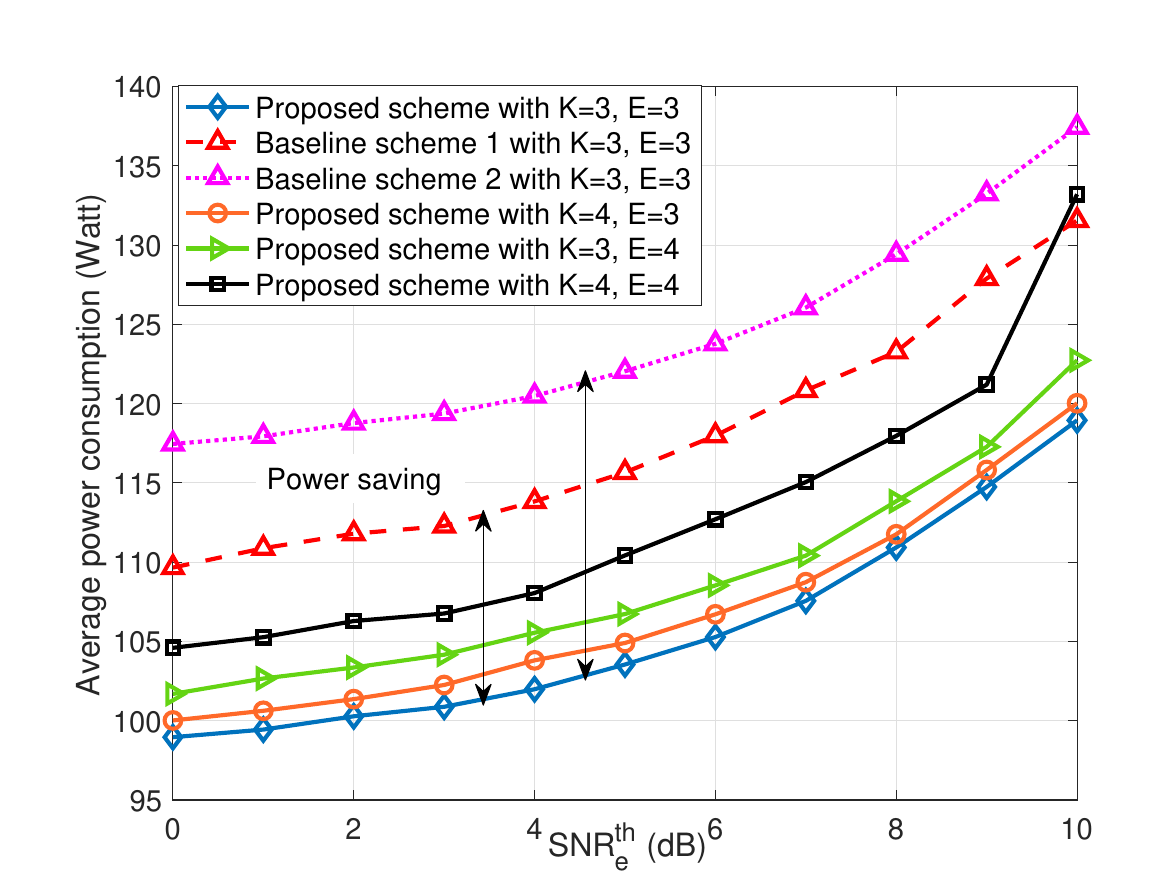}
\caption{Average power consumption versus the minimum required sensing SNR.}
\end{figure}
\begin{figure}[t]
 \centering
		\label{figure5b}
		\includegraphics[width=8.6cm]{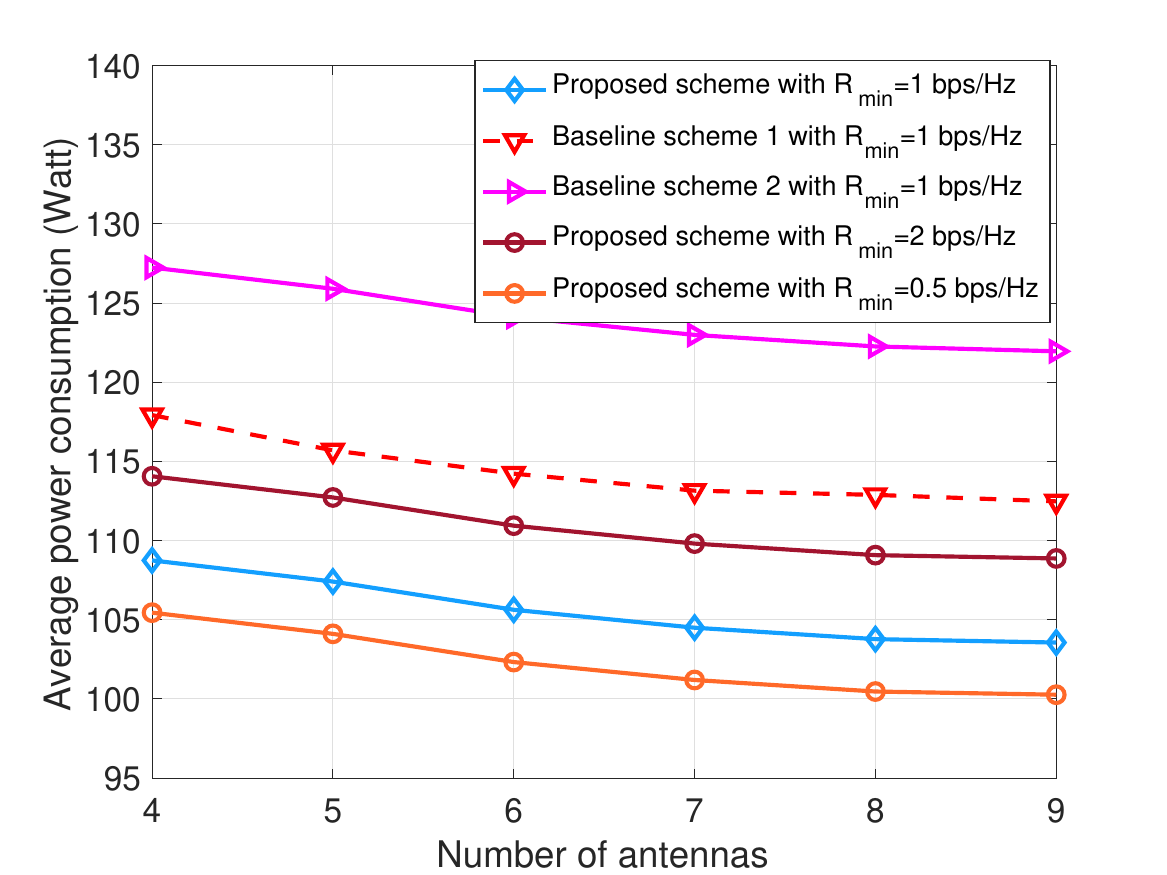}
  \caption{Average power versus the number of antennas.}
  \end{figure}
  \begin{figure}
   \centering
  \includegraphics[width=8.6cm]{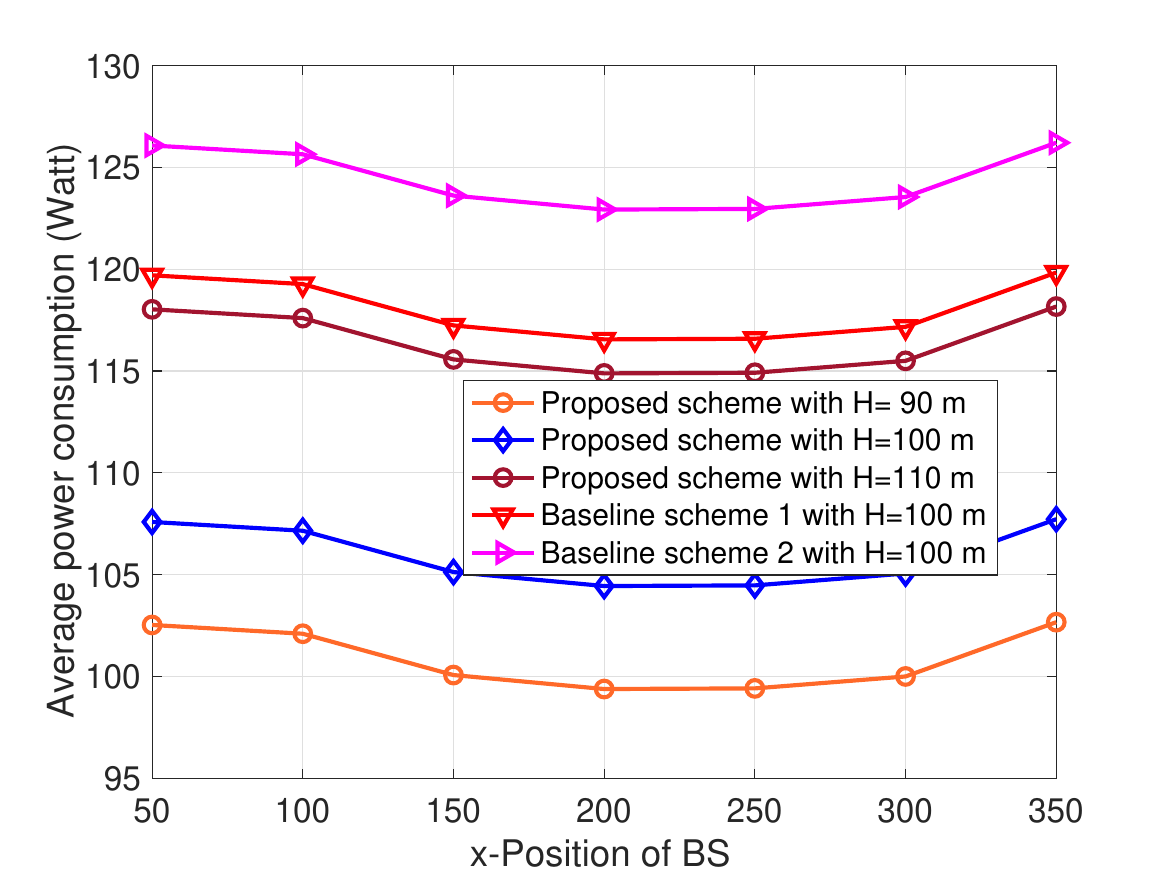}
      \caption{Average transmit power versus position of BS on the x-axis for various UAV heights.}
      \label{fig:enter-label}
  \end{figure}

Fig. 5 shows the average power consumption versus the sensing SNR requirement.~The average power consumption of the UAV for both the proposed scheme and the baseline schemes increases monotonically with the minimum SNR threshold for sensing. This escalation occurs because, in order to meet more stringent sensing requirements, the UAV must not only transmit with higher power but also dedicate more time to hovering. Furthermore, Fig. 5 highlights the longer impact of the number of sensing targets on power consumption compared to the number of communication users. This is primarily attributed to two factors. Firstly, sensing tasks necessitate higher transmission power because of the round-trip pathloss, thereby increasing power consumption. Secondly, our proposed scheme requires UAV hovering during sensing, which consumes more power compared to the flight mode. These factors collectively highlight the significant role of the sensing tasks on power consumption. Nevertheless, the average power consumption does increase with the number of communication users. This increase can be attributed to the UAV-mounted transmitter needing to allocate more degrees of freedom (DoFs) to mitigate multi-user interference (MUI). However, this diminishes the flexibility in trajectory and beamforming design, ultimately resulting in performance degradation. The impact of velocity and trajectory optimization on UAV power consumption is also evident in Fig. 5. Specifically, the proposed scheme, leveraging trajectory design to provide additional DoFs, consumes less power compared to baseline scheme 1 with a fixed trajectory. Likewise, baseline scheme 2 incurs a higher power consumption due to both a fixed beamforming policy, leading to increased transmit power, and a fixed UAV velocity leading to an increased aerodynamic power consumption.

Fig. 6 shows the average power consumption versus the number of antennas at the UAV for different minimum data rate requirements, denoted as $R_{\min}$. As can be seen, the average UAV power consumption decreases with increasing number of transmit antennas. This is because the extra DoFs offered by the additional antennas facilitate more precise beamforming and can efficiently mitigate MUI. However, as the number of antennas increases, the performance gains decreases, suggesting reduced marginal benefits. In this context, it is essential to note the impact of the circuit power consumption. While additional antennas can improve beamforming, the resulting reduction in power consumption are counteracted by the additional circuit power required. This trade-off suggests that, beyond a certain point, the increase in circuit power consumption may outweigh the beamforming gains achieved with additional antennas. We also observe a notable difference in power consumption between the proposed scheme and the two baseline schemes. Specifically, in baseline scheme 2, the fixed information beamforming policy leads to an increased transmit power. This is because the scheme fails to fully exploit the spatial DoFs. Moreover, a substantial amount of power is consumed during the UAV's flight in baseline scheme 2, which is attributed to the fixed velocity of the UAV, resulting in elevated aerodynamic power consumption. As a consequence, baseline scheme 2 exhibits higher overall power consumption compared to the proposed scheme. Additionally, Fig. 6 reveals that the average power consumption increases as the minimum quality of service requirement ($R_{\min}$) becomes more stringent. This is because to satisfy a stricter minimum required data rate, the UAV needs to increase its transmit power, resulting in higher power consumption.

Fig. 7 shows the impact of the position of the BS on the average power consumption, considering different heights of the UAV. In particular, the BS is fixed on the y-axis at $y=0$ m, while it is moved along the x-axis from $x=50$ m to $x=350$ m. As can be observed, the average power consumption of the UAV decreases as the BS moves along the x-axis until it reaches $x=250$~m. This reduction is attributed to the closer proximity of the BS to the first and second sensing targets, requiring less power for data offloading when the UAV visits these targets. Notably, at $x=200$ m and $x=250$ m, where the BS is close to the first and second targets, respectively, the UAV can transmit with lower power during offloading. Conversely, for $x>250$ m, where the distance to the first and second targets is higher, more transmit power is required to ensure successful data offloading. Furthermore, the proposed algorithm demonstrates superior power efficiency compared to both baseline schemes. Additionally, we observe from Fig. 7 that as the UAV's altitude increases, two key effects come into play: the vertical distance to the targets increases, demanding more power for precise sensing, and the distance to both the communication users and the BS also grows, requiring a higher transmit power to meet the QoS constraints of the users and successful offloading.
\begin{figure}[t]
 \centering
		\label{figure9}
		\includegraphics[width=8.6cm]{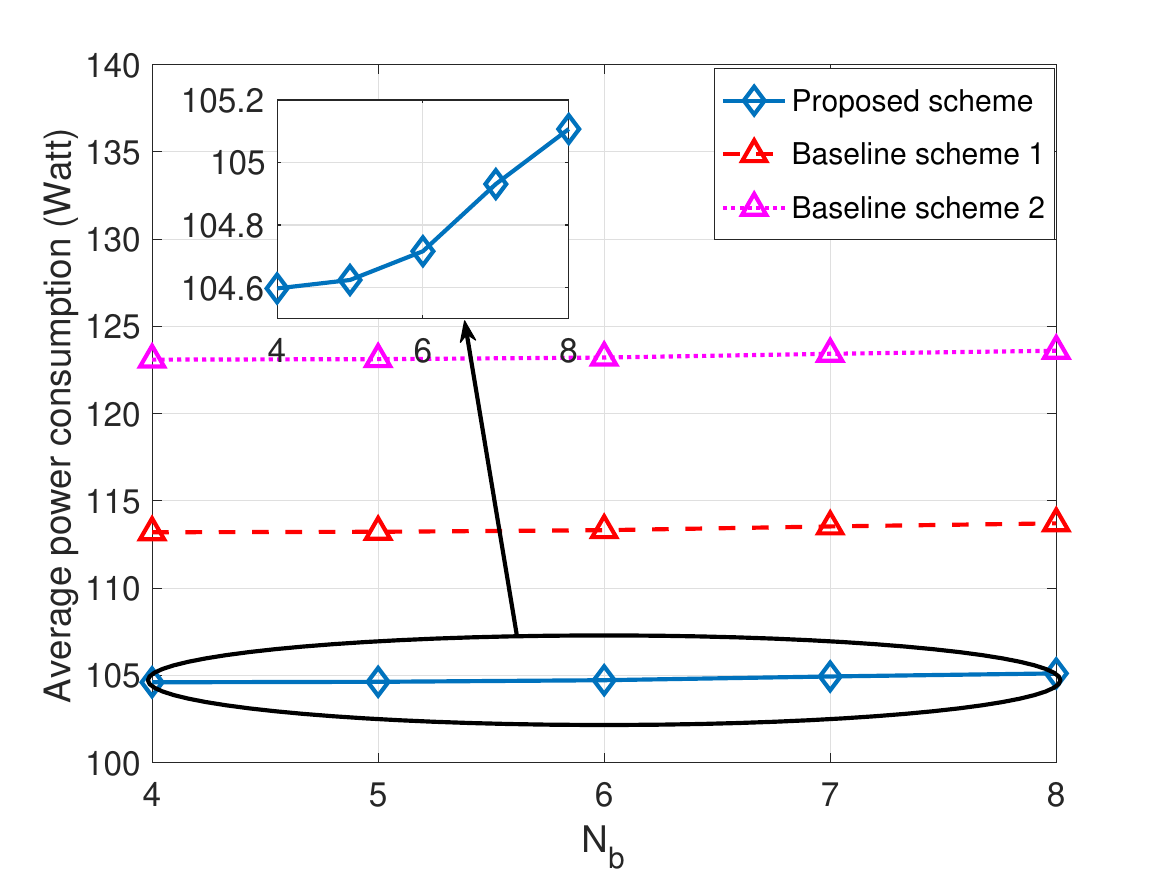}
  \caption{Average power consumption versus number of bits $N_{b}$.}
  \end{figure}
  
Fig. 8 reveals that the average power consumption is monotonically non-decreasing with respect to the number of bits $N_{b}$ used for quantizing echoes. This is because as $N_{b}$ increases, the rate of production $R_{\text{Pr}}$ in \eqref{rate_production} also grows, requiring additional power expenditure for data offloading to meet constraint C4, underlining the impact of quantization resolution on both energy efficiency and operational costs. The relationship between bit resolution, data volume, and power consumption is crucial, especially in scenarios where data offloading becomes a bottleneck for system performance.
\begin{figure}[t]
 \centering
		\label{figure9}
		\includegraphics[width=8.6cm]{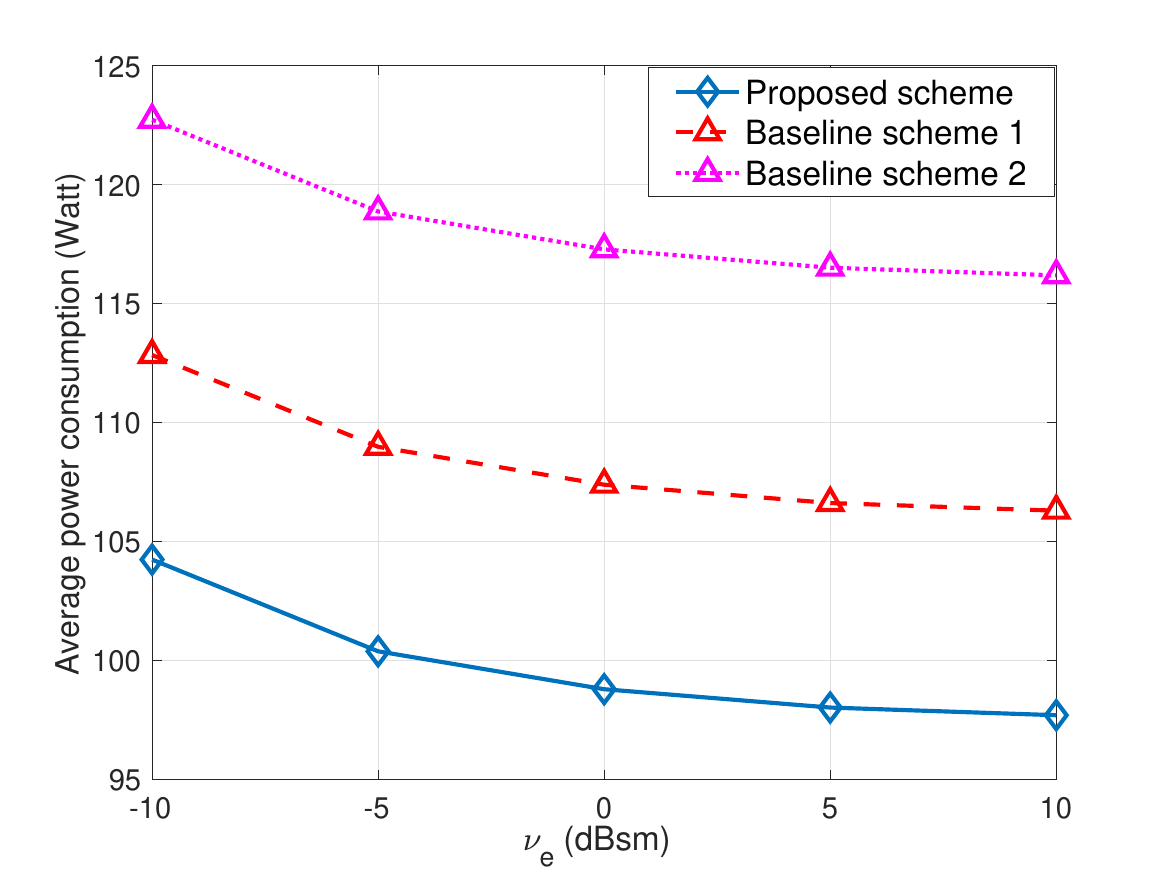}
  \caption{Average power consumption versus RCSs.}
  \end{figure}
  
 Fig. 9 depicts the impact of the RCS on the average power consumption for the proposed scheme and the baseline schemes. The RCS is expressed in decibels relative to one square meter (dBsm)\cite{Richards2010}. As the RCS increases, there is a notable decrease in average power consumption for all considered schemes. This inverse relationship is attributed to the enhanced radar reflectivity provided by larger RCSs, which allows the UAV to reduce the radar transmit power. The reduction in transmit power is possible because the enhanced reflectivity ensures that the target remains detectable, as the required sensing SNR threshold can be met with less energy expenditure.
 \section{Conclusion and Future Work}
In this paper, we studied the joint resource allocation and
trajectory design for a multi-user, multi-target UAV-based ISAC
system, where we accounted the limited capacity of the backhaul link, which is needed for offloading the sensing data to the BS. 
To avoid interference between sensing and communication, sensing and communication were performed in orthogonal time slots, where the time of sensing was optimized. To be compatible with practical UAV-based sensing systems, pulse radar-based sensing was carried out during UAV hovering. Taking into account the application of a focused sensing beam with small sidelobes, a minimum required accumulated sensing SNR, and UAV hovering during sensing, we minimized the average power consumption of the UAV while ensuring the QoS for both the communication users and the sensing tasks. To solve the resulting challenging non-convex MINLP, we developed a computationally efficient AO-based algorithm, which yielded a high-quality suboptimal solution. Our simulation results revealed that 1) the proposed
design enables substantial power savings compared to two
baseline schemes; 2) more stringent sensing requirements lead
to longer sensing times, highlighting the trade-off between
sensing accuracy and sensing time; 3) the number of sensing targets has a larger impact on power consumption than the
number of communication users; 4) larger RCSs enhance the echo signal strength, thereby facilitating a reduction in radar transmit power without compromising sensing performance; 5) the average power consumption increases with the number of bits $N_{b}$ used for quantizing echoes; 6) data offloading can be
improved by positioning the BS closer to the sensing targets;
7) the optimized trajectory design ensures precise hovering
above the target during sensing, facilitating high-quality sensing
with energy-focused beams; and 8) the designed UAV trajectory balances the distance of the UAV to the communication users and the BS.

An interesting topic for future research on UAV-enabled ISAC systems is the investigation of multi-UAV sensing strategies. The deployment of multiple UAVs might enable simultaneous sensing of multiple targets, expanding operational capabilities and enhancing the flexibility in sensing. However, the related challenges, such as interference management, trajectory optimization, and collision avoidance, have to be carefully addressed. Additionally, adaptive beam pattern designs that enable sensing from multiple angles around a target could further enhance performance and flexibility. Furthermore, advanced dynamic beam steering techniques may be investigated to adapt UAV sensing in real-time to environmental changes and moving targets at the expense of a higher system complexity. 
\bibliography{Mybib}
\bibliographystyle{IEEEtran}
\begin{IEEEbiography}[{\includegraphics[width=1.25in,height=1.25in,clip,keepaspectratio]{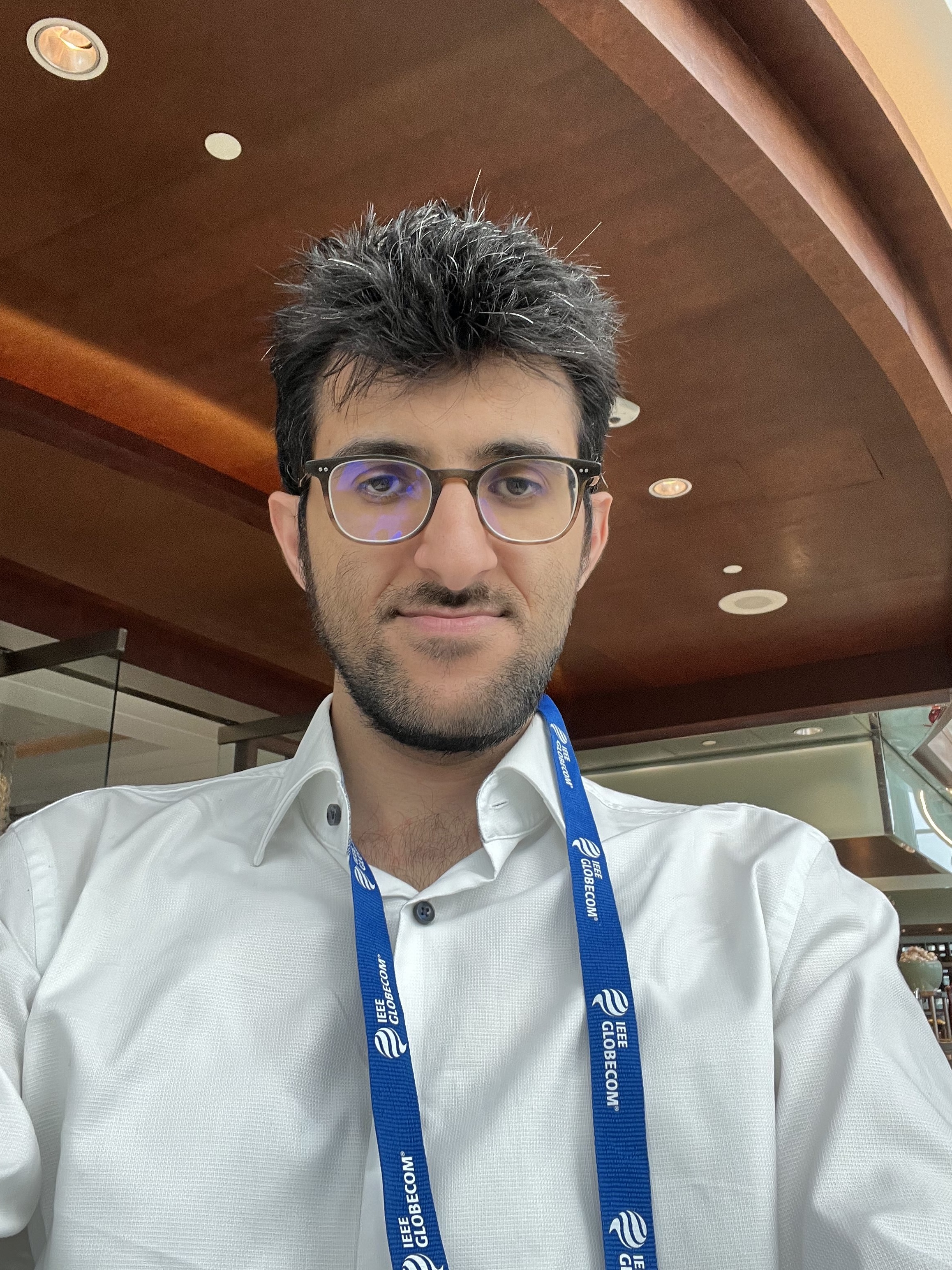}}]{Ata Khalili (Member, IEEE)}
received the B.Sc. and M.Sc. degrees (Hons.) in electronic engineering and telecommunication engineering from Shahed University in 2016 and 2018, respectively. He is currently pursuing the Ph.D. degree with the Institute for Digital Communications, Friedrich-Alexander University of Erlangen–Nürnberg, Erlangen, Germany. He has been serving as a member of the Technical Program Committees for the \textsc{IEEE Globecom}, \textsc{IEEE WCNC}, and \textsc{IEEE ICC} conferences, since 2020. He served as a Session Chair for IEEE GLOBECOM 2021 and IEEE ICC 2022. He is a Reviewer of several IEEE journals, such as IEEE \textsc{Journal on Selected Areas in Communications} and \textsc{IEEE Transactions on Wireless Communications}. He received the Best Paper Award at IEEE WPMC 2022 and was recognized as an Exemplary Reviewer for the \textsc{IEEE Transactions on Communications} and the \textsc{IEEE Wireless Communications Letters} in 2022.
\end{IEEEbiography}
\begin{IEEEbiography}
    {Atefeh Rezaei} (Student Member, IEEE) received the M.Sc. degree in electrical engineering from Tarbiat Modares University (TMU), Tehran, Iran, in 2016. She is currently pursuing the Ph.D. degree at the Technical University of Berlin, Berlin, Germany. 
    Her current research interests include the design and analysis of wireless communication networks emphasizing the application of optimization theory, MIMO radar, signal processing, and machine learning.
\end{IEEEbiography}
\begin{IEEEbiography}
[{\includegraphics[width=0.95in,height=1.25in]{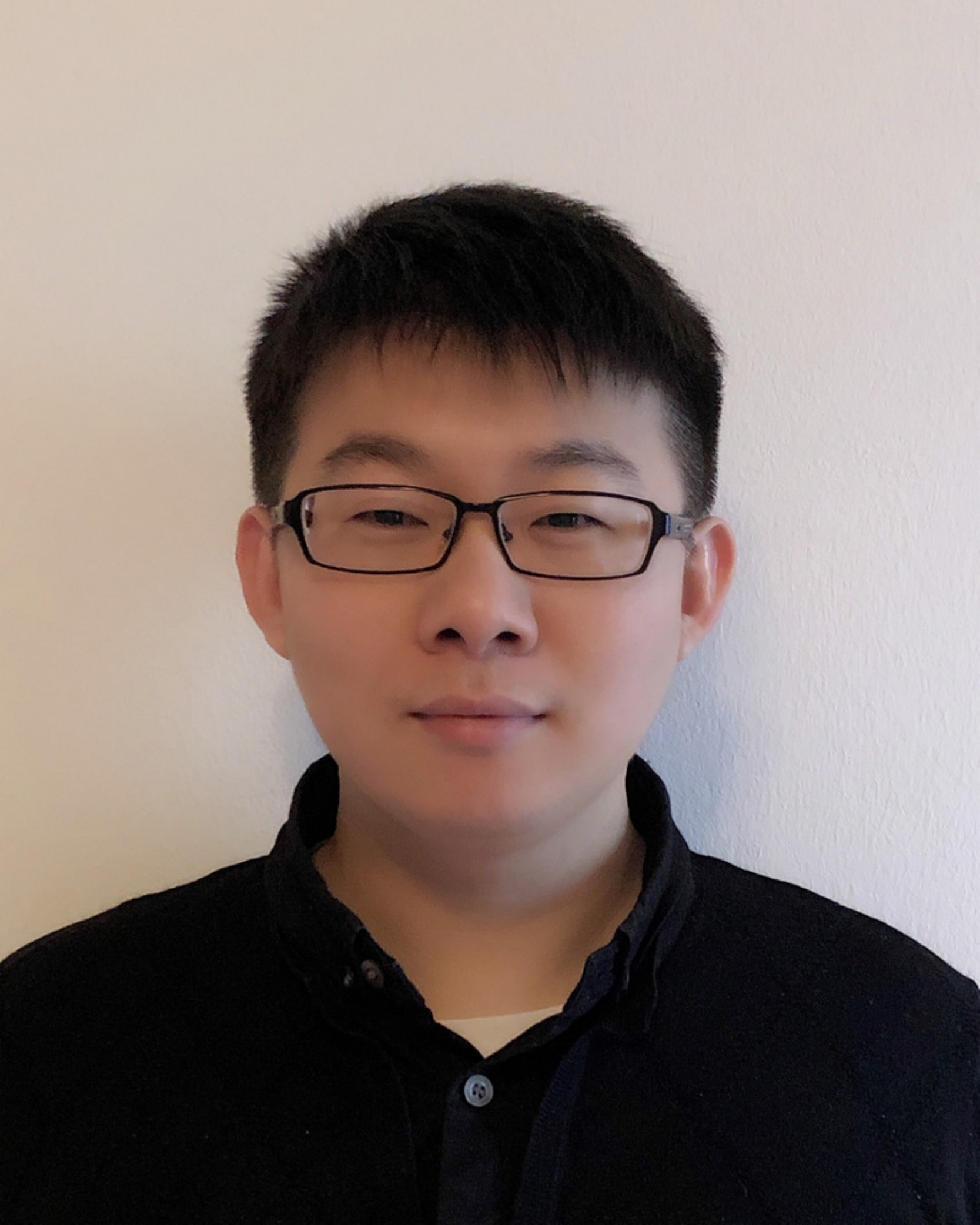}}]{Dongfang Xu} (Member, IEEE) received his Ph.D. degree (with distinction)
in electrical engineering from Friedrich-Alexander-Universität Erlangen-Nürnberg
(FAU), Erlangen, Germany in December 2022. Previously, he received his M.Sc. degree (with distinction) in communications and multimedia engineering from FAU in September 2017 and his B.Eng. degree in communication engineering from Shandong University, Jinan, China in July 2014, respectively.
\par
Currently, he is a research assistant professor at the Hong Kong University of Science and Technology (HKUST). He was a co-recipient of the IEEE Communications Society Leonard G. Abraham Prize 2023, the IEEE Communications Society Stephen O. Rice Prize 2022, and the IEEE Global Communications Conference (GLOBECOM) 2019 Best Paper Award. He was also recognized as an Exemplary Reviewer of the \textsc{IEEE Transactions on Communications} in 2020, 2021, and 2022 and an Exemplary Reviewer of the \textsc{IEEE Communications Letters} in 2023.
\end{IEEEbiography}
\begin{IEEEbiography}
[{\includegraphics[width=1.1in,height=1.25in]{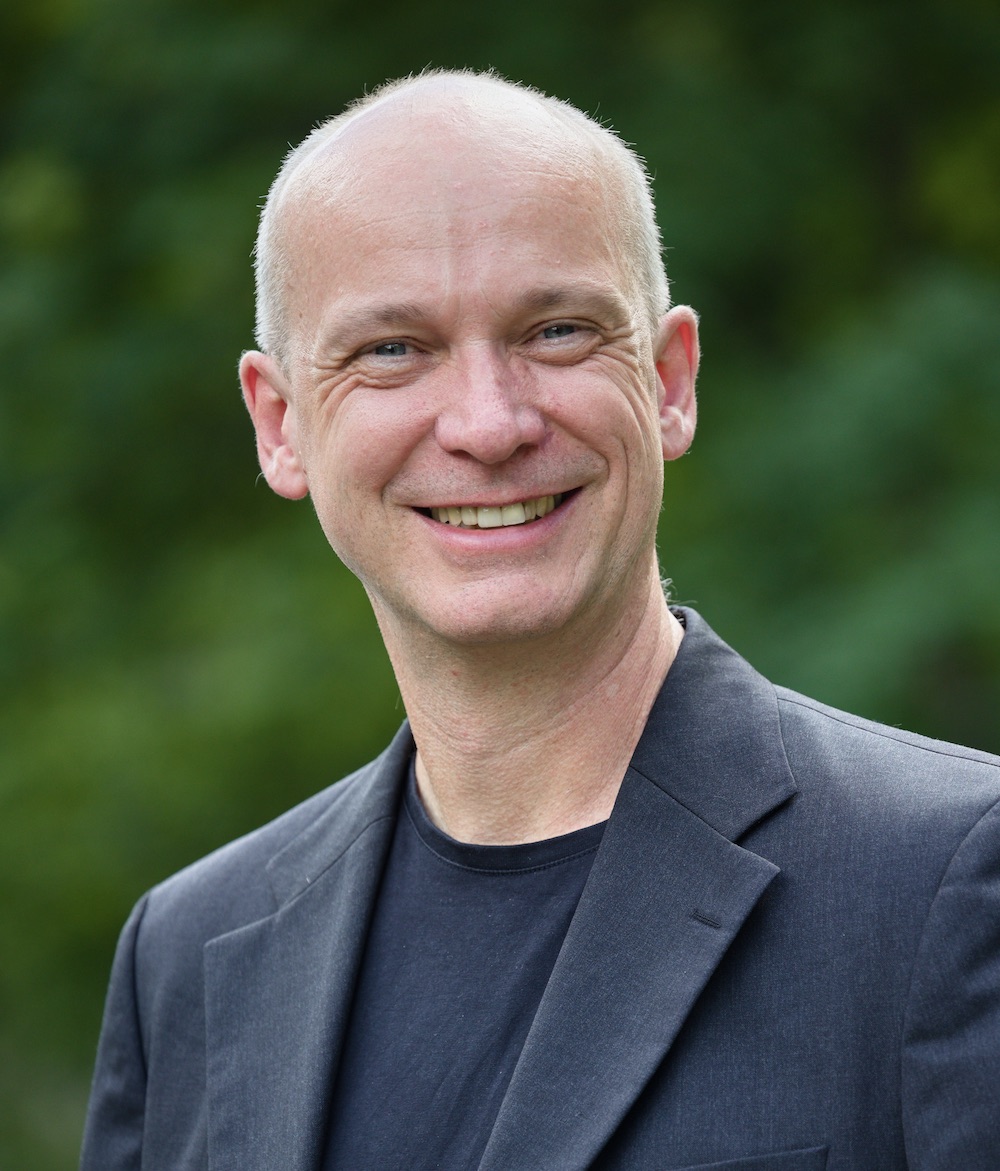}}]
    {Falko Dressler} is full professor and Chair for Telecommunication Networks at the School of Electrical Engineering and Computer Science, TU Berlin. He received his M.Sc. and Ph.D. degrees from the Dept. of Computer Science, University of Erlangen in 1998 and 2003, respectively. Dr. Dressler has been associate editor-in-chief for IEEE Trans. on Network Science and Engineering, IEEE Trans. on Mobile Computing and Elsevier Computer Communications as well as an editor for journals such as IEEE/ACM Trans. on Networking, Elsevier Ad Hoc Networks, and Elsevier Nano Communication Networks. He has been chairing conferences such as IEEE INFOCOM, ACM MobiSys, ACM MobiHoc, IEEE VNC, IEEE GLOBECOM. He authored the textbooks Self-Organization in Sensor and Actor Networks published by Wiley \& Sons and Vehicular Networking published by Cambridge University Press. He has been an IEEE Distinguished Lecturer as well as an ACM Distinguished Speaker. Dr. Dressler is an IEEE Fellow as well as an ACM Distinguished Member. He is a member of the German National Academy of Science and Engineering (acatech). He has been serving on the IEEE COMSOC Conference Council and the ACM SIGMOBILE Executive Committee. His research objectives include next generation wireless communication systems in combination with distributed algorithms for applications in the Internet of Things, and Cyber-Physical Systems, and the Internet of Bio-Nano-Things.
\end{IEEEbiography}
\begin{IEEEbiography}
[{\includegraphics[width=0.95in,height=1.25in]{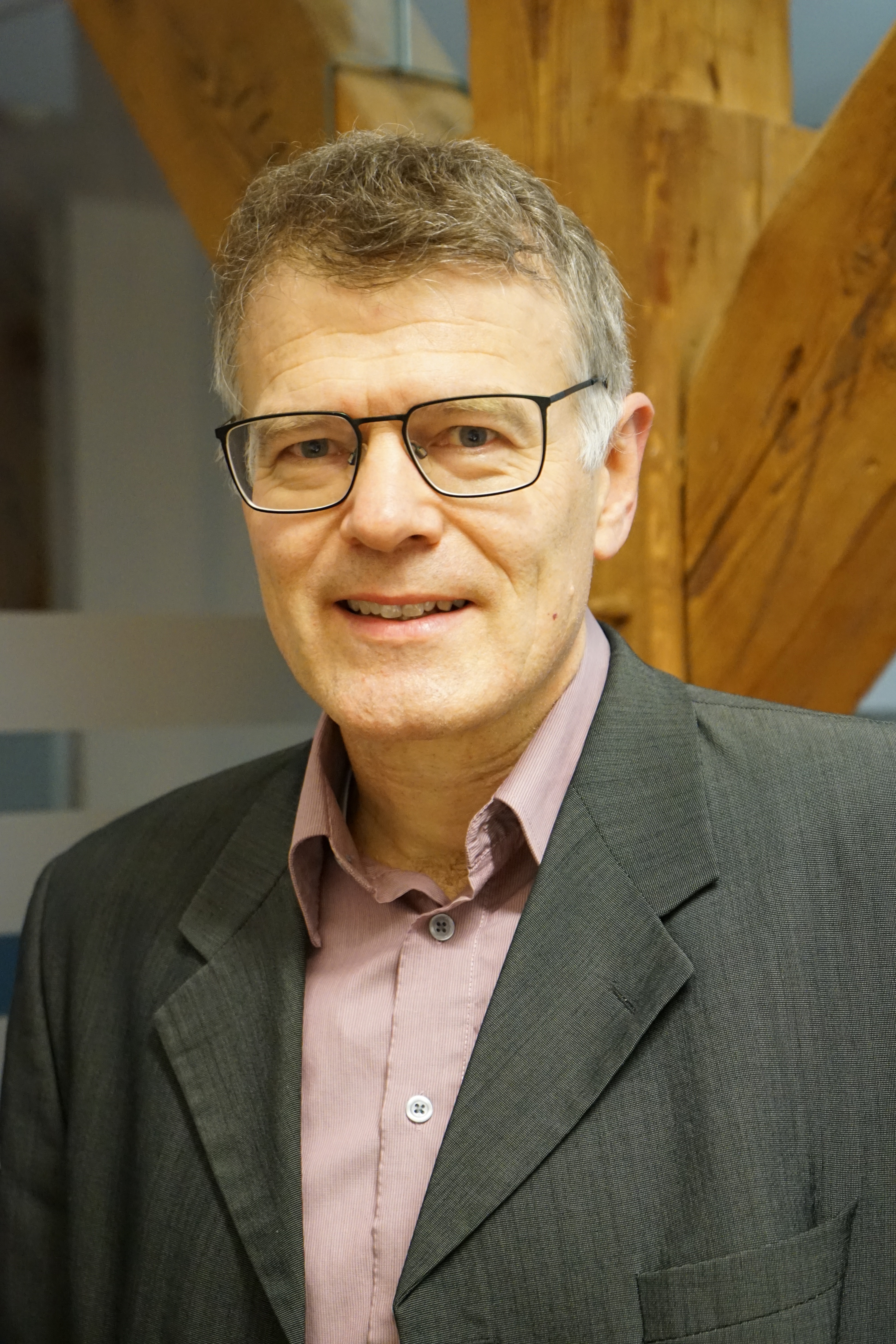}}]
    {Robert Schober} (S'98, M'01, SM'08, F'10) received the Diplom (Univ.) and the Ph.D. degrees in electrical engineering from Friedrich-Alexander University of Erlangen-Nuremberg (FAU), Germany, in 1997 and 2000, respectively. From 2002 to 2011, he was a Professor and Canada Research Chair at the University of British Columbia (UBC), Vancouver, Canada. Since January 2012 he is an Alexander von Humboldt Professor and the Chair for Digital Communication at FAU. His research interests fall into the broad areas of Communication Theory, Wireless and Molecular Communications, and Statistical Signal Processing.
 \par
Robert received several awards for his work including the 2002 Heinz Maier Leibnitz Award of the German Science Foundation (DFG), the 2004 Innovations Award of the Vodafone Foundation for Research in Mobile Communications, a 2006 UBC Killam Research Prize, a 2007 Wilhelm Friedrich Bessel Research Award of the Alexander von Humboldt Foundation, the 2008 Charles McDowell Award for Excellence in Research from UBC, a 2011 Alexander von Humboldt Professorship, a 2012 NSERC E.W.R. Stacie Fellowship, a 2017 Wireless Communications Recognition Award by the IEEE Wireless Communications Technical Committee, and the 2022 IEEE Vehicular Technology Society Stuart F. Meyer Memorial Award. Furthermore, he received numerous Best Paper Awards for his work including the 2022 ComSoc Stephen O. Rice Prize and the 2023 ComSoc Leonard G. Abraham Prize. Since 2017, he has been listed as a Highly Cited Researcher by the Web of Science. Robert is a Fellow of the Canadian Academy of Engineering, a Fellow of the Engineering Institute of Canada, and a Member of the German National Academy of Science and Engineering. He served as Editor-in-Chief of the IEEE Transactions on Communications, VP Publications of the IEEE Communication Society (ComSoc), ComSoc Member at Large, and ComSoc Treasurer. Currently, he serves as Senior Editor of the Proceedings of the IEEE and as ComSoc President.
 
\end{IEEEbiography}
\end{document}